\documentclass[fleqn,usenatbib]{mnras}

\usepackage[T1]{fontenc}
\usepackage{ae,aecompl}

\usepackage{graphicx}	
\usepackage{amsmath}	
\usepackage{amssymb}	
\usepackage{amstext}
\usepackage{caption}

\usepackage{mathtools}
\usepackage{siunitx}
\usepackage{physics}
\usepackage{xspace}
\usepackage{cleveref}
\usepackage{pdflscape}
\usepackage[online]{threeparttable}

\usepackage{newtxtext,newtxmath}

\DeclareSIUnit{\parsec}{pc}
\DeclareSIUnit{\pc}{pc}
\DeclareSIUnit{\year}{yr}
\newcommand{\Msun}{\ensuremath{\mathrm{M}_{\odot}}}

\newcommand{\limepy}{\textsc{limepy}\xspace}
\newcommand{\Nbody}{\(N\)-body\xspace}

\newcommand*\NGC[1]{NGC\thinspace{#1}}
\newcommand*\Messier[1]{M\thinspace{#1}}
\newcommand{\omegacen}{\(\omega\)\thinspace{Cen}\xspace}

\newcommand*\chem[1]{\ensuremath{\mathrm{#1}}}
\newcommand{\FeH}{\ensuremath{[\chem{Fe}/\chem{H}]}}

\newcommand{\ra}{\ensuremath{r_{\mathrm{a}}}}
\newcommand{\raj}{\ensuremath{r_{\mathrm{a},j}}}
\newcommand{\rh}{\ensuremath{r_{\mathrm{h}}}}
\newcommand{\rt}{\ensuremath{r_{\mathrm{t}}}}
\newcommand{\BHret}{\ensuremath{\mathrm{BH}}_{\mathrm{ret}}}

\newcommand{\paperII}{Paper~II}

\title[Multimass modelling of GCs - I. High-mass IMF]{
	Multimass modelling of Milky Way globular clusters
	- I. Implications on their stellar initial mass function above \(1\ \Msun\)
}

\author[N. Dickson et al.]{
N. Dickson$^{1}$\thanks{E-mail: nolan.dickson@smu.ca}, 
V. Hénault-Brunet$^{1}$, H. Baumgardt$^{2}$, M. Gieles$^{3,4}$,
P.J. Smith$^{1}$ 
\\
$^{1}$Department of Astronomy and Physics, Saint Mary's University,
923 Robie Street, Halifax, NS B3H 3C3, Canada \\
$^{2}$School of Mathematics and Physics, The University of Queensland, St Lucia, QLD 4072, Australia\\
$^{3}$ICREA, Pg. Llu\'{i}s Companys 23, E08010 Barcelona, Spain\\
$^{4}$Institut de Ci\`{e}ncies del Cosmos (ICCUB), Universitat de Barcelona (IEEC-UB), Mart\'{i} Franqu\`{e}s 1, E08028 Barcelona, Spain
}

\date{Accepted XXX. Received YYY; in original form ZZZ}

\pubyear{2023}

\begin{document}
\label{firstpage}
\pagerange{\pageref{firstpage}--\pageref{lastpage}}
\maketitle

\begin{abstract}

The distribution of stars and stellar remnants (white dwarfs, neutron stars,
black holes) within globular clusters holds clues about their formation and
long-term evolution, with important implications for their initial mass
function (IMF)
and the formation of black hole mergers. In this work, we present
best-fitting multimass models for 37 Milky Way
globular clusters, which were inferred from various datasets, including proper
motions from Gaia EDR3 and HST, line-of-sight velocities from ground-based
spectroscopy and deep stellar mass functions from HST.
We use metallicity dependent stellar evolution recipes to obtain present-day mass
functions of stars and remnants from the IMF.
By dynamically probing the present-day mass function of all objects in a cluster,
including the mass distribution of remnants, these models allow us to explore
in detail the stellar (initial) mass functions of a large sample of Milky Way GCs.
We show that, while the low-mass mass function slopes are strongly dependent on
the dynamical age of the clusters, the high-mass slope
(\(\alpha_3; m > \SI{1}{\Msun}\)) is not, indicating that the mass function in
this regime has generally been less affected by dynamical mass loss.
Examination of this high-mass mass function slope
suggests an IMF in this mass regime consistent with a Salpeter IMF is required
to reproduce the observations. This high-mass IMF is incompatible with a
top-heavy IMF, as has been proposed recently.
Finally, based on multimass model fits to our sample of
Milky Way GCs, no significant correlation is found between the
high-mass IMF slope and cluster metallicity.

\end{abstract}





\section{Introduction}\label{sec:introduction}


    The stellar initial mass function (IMF) plays a key role in the
    evolutionary history and properties of populations of stars,
    and understanding it is vital to understanding and interpreting both
    observations and simulations of star clusters and galaxies.

    Globular clusters (GCs) consist of very large numbers of stars of
    similar iron abundance and age, providing us with
    one of the best avenues for investigating the shape of the stellar IMF,
    and how it may vary with environment.
    The IMF plays a particularly important role in the evolution of GCs,
    where it controls the populations of stellar remnants, the degree and
    timescale of mass segregation, the lifetime of the clusters before
    dissolution, and the contribution of GCs to observed gravitational
    waves \citep[e.g.][]{Haghi2020,weatherford2021,Wang2021}.


    The universality of the stellar IMF is a debated topic.
    While the typically assumed (canonical) formulations of the IMF, determined
    empirically through observations of solar-neighbourhood and Milky Way
    cluster stars \citep[e.g.][]{Salpeter1955,Kroupa2001,Chabrier2003}, seem to
    demonstrate that it is universal among star-forming systems, the
    exact shape and universality of the IMF is still under investigation
    \citep[see review by][]{Bastian2010}.
    For example, observations of the cores of early-type galaxies,
    (both spectroscopic; \citealp{vanDokkum2010} and kinematic;
    \citealp{Cappellari2012}) have pointed towards a ``bottom-heavy'' IMF,
    enriched with low-mass stars, in those environments (although see also
    \citealp{Smith2014,Smith2020}).
    Meanwhile, recent theoretical studies of star and cluster formation have
    indicated that certain environment-dependent processes, such as radiative
    feedback or cooling from dust-grains, could imply a varying IMF
    \citep{Krumholz2011,Chon2021}.


    In particular for GCs, some recent observational works have also showcased
    trends which could be explained by a varying IMF. These
    observations, however, have also been shown to be explainable without the
    need to invoke such a non-canonical IMF.
    \citet{strader2011} demonstrated that dynamical mass measurements of 200
    globular clusters in \Messier{31} showed a decreasing trend in the dynamical
    mass-to-light ratio with increasing cluster metallicity. This result
    is opposite to what standard stellar population models would predict
    while assuming a canonical IMF. \citet{Haghi2017} showed that these
    results could be explained by introducing a non-canonical,
    metallicity-dependent IMF, with an increasing level of top-heaviness for
    low metallicity clusters \citep{Marks2013}. 
    %
    However \citet{Baumgardt2020}, in a study of Milky Way GCs,
    also noted that such a discrepancy in the mass-to-light ratios compared to
    population synthesis models could be accounted for once the low-mass
    depleted present-day mass function (PDMF) of the metal-rich clusters was
    taken into consideration. Metallicity-dependent stellar evolution models
    were also able to account for the difference in the metal-poor clusters.
    \citet{Shanahan2015} also demonstrated that not accounting for
    mass segregation in integrated-light studies of \Messier{31} clusters
    introduces a bias in the inferred dynamical mass,
    dependent on metallicity \citep[see also][]{Sippel2012},
    and argued that there is no need for variations in the IMF to explain the
    \citeauthor{strader2011} results.
    Because the majority of stars form in GCs at low metallicities
    \citep{Larsen2012}, a substantially flatter IMF at high masses in
    GCs would have important consequences for the amount of ionization radiation
    at high redshift \citep{Schaerer2011,BoylanKolchin2018}; the chemical
    evolution of galaxies; the amount of
    stellar-mass black holes (BHs) formed and subsequent binary BH mergers
    \citep{Schneider2018}. A flat IMF also predicts that there should be
    more white dwarfs (WDs) at the present age. WDs contribute \(\sim30\%\) to
    the total mass at \(\sim 10\,\mathrm{Gyr}\) for a canonical IMF, before
    accounting for the preferential loss of low-mass objects in GCs. The
    fractional contribution to the central density is higher because of mass
    segregation, so it is feasible to look for an excess of WDs in the
    kinematics of GCs.


    In this work, the hypothesis of metallicity-dependent, variable and
    non-canonical stellar IMFs in globular clusters is investigated,
    in particular in the high-mass regime, where stars in old globular clusters
    have, by the present day, evolved into stellar remnants.
    To do so, we fit multimass dynamical models to various observables for
    a large sample of Milky Way clusters, over a range of metallicities.
    We infer their global stellar mass functions
    and simultaneously constrain their distributions of stellar remnants.
    The multimass \limepy models and mass function evolution algorithm used are
    explained in more detail in \Cref{sec:models}. \Cref{sec:cluster_data}
    describes the methods and sources used to obtain all observational data
    used to fit the models, as well as how the cluster sample was chosen. The
    model fitting procedure, including descriptions of all probability distributions and
    Bayesian sampling techniques, as well as the software library and fitting
    pipeline which was created to facilitate this fitting, is presented in
    \Cref{ch:model_fitting}.
    The results of the fitting of all clusters in our sample based on these
    methods are given in \Cref{ch:results}. The (initial) mass function results
    for all clusters are presented and explored in more detail in
    \Cref{ch:mass_functions}.
    Finally, we conclude in \Cref{ch:conclusions}.

    The inferred present-day populations of stellar-mass BHs in
    our sample of globular clusters based on our best-fitting models
    will be examined in detail in a separate
    paper (Dickson et al., in prep.; hereafter \paperII).



\section{Models}\label{sec:models}

    To model the mass distribution of the globular clusters analyzed in this
    work, we use the \limepy multimass distribution-function (DF) based models
    \citep{Gieles2015}
    \footnote{Available at \url{https://github.com/mgieles/limepy}}.
    DF based models are equilibrium models built around a distribution
    function \(f\) which describes the particle density of stars and satisfies
    the collisionless-Boltzmann equation.
    This DF is used to self-consistently solve for the system's potential
    (\(\phi(r)\)) using Poisson's equation.

    A variety of quantities can be derived from the DF which can be used to
    describe a globular cluster, including the projected velocity dispersion
    (the second velocity moment), the projected surface density, the total mass,
    the potential energy and the system entropy
    \citep[e.g.][]{Spitzer1987,Gieles2015}. Observational data can be used
    to compare and constrain the models based on these quantities.

    Multimass models allow for a more accurate description of real globular
    clusters, which are made up of a spectrum of stellar masses.
    Multiple mass components are necessary in order to describe the
    distributions of different stellar and remnant populations within the
    system and, in turn, examine both the process and effects of mass
    segregation \citep[e.g.][]{Dacosta1976}.

    The DF of the multimass version of the \limepy models is given by the sum
    of component DFs for every mass bin \(j\), each as a function of
    the specific energy \(E\) and angular momentum \(J\) in the form of:
    \begin{equation}\label{eq:DF_limepy}
        f_{j}(E, J^2) =
            \begin{dcases}
                A_j\, \exp\left(-\frac{J^2}{2 \raj^2 s_j^2}\right)\,
                E_g \left(-\frac{E-\phi(\rt)}{s_j^2}\right) & E < \phi(r_t), \\
                0 & E \geq \phi(r_t),
            \end{dcases}
    \end{equation}
    where \(A_j\) and \(s_j\) are the mass-dependant normalization
    and velocity scales, \raj and \rt are the anisotropy and
    truncation radii, and the function \(E_g\) is defined using the regularized
    lower incomplete gamma function and the truncation parameter \(g\):
    \begin{equation}
        E_g(x) =
        \begin{dcases}
            \exp(x) & g=0, \\
            \exp(x) \frac{\gamma(g,x)}{\Gamma(g)} & g>0,
        \end{dcases}
    \end{equation}
    These parameters and how they are used are explained in more detail
    in \Cref{sub:model_parameters} below.

\subsection{Model parameters}\label{sub:model_parameters}

    Our models are defined by 10 free parameters (listed in Table
    \ref{table:free_params}) which dictate the mass function and physical
    solution of the \limepy DF.
    
    \begin{table*}
    \renewcommand*{\arraystretch}{1.4}
    \centering
    \begin{tabular}{ c l l }
        \hline
        Parameter & Description & Prior \\
        \hline
        \(\hat{\phi}_0\) & Dimensionless central potential & \(\mathrm{Uniform}(L=2.0,\ U=15.0)\) \\
        \(M\) & Total system mass \(\left[10^6\ \Msun\right]\) & \(\mathrm{Uniform}(L=0.001,\ U=2.0)\) \\
        \(g\) & Truncation parameter & \(\mathrm{Uniform}(L=0.0,\ U=3.5)^\ast\) \\
        \(\rh\) & Half-mass radius \([\mathrm{pc}]\) & \(\mathrm{Uniform}(L=0.5,\ U=15.0)\) \\
        \(\log(\hat{r}_{\mathrm{a}})\) & Dimensionless anisotropy radius & \(\mathrm{Uniform}(L=0.0,\ U=8.0)\) \\
        \(\delta\) & Velocity-scale mass dependence & \(\mathrm{Uniform}(L=0.3,\ U=0.5)^\ast\) \\
        \hline
        \(\alpha_1\) & MF exponent \((0.1\ \Msun<m\leq 0.5\ \Msun)\) & \(\mathrm{Uniform}(L=-1.0,\ U=2.35)^\ast\) \\
        \(\alpha_2\) & MF exponent \((0.5\ \Msun<m\leq 1\ \Msun)\) & \(\mathrm{Uniform}(L=-1.0,\ U=\mathrm{min}(2.35,\ \alpha_1))^\ast\) \\
        \(\alpha_3\) & MF exponent \((1\ \Msun<m\leq 100\ \Msun)\) & \(\mathrm{Uniform}(L=1.6,\ U=\mathrm{min}(4.0,\ \alpha_2))^\ast\) \\
        \(\BHret\) & Black hole retention fraction \([\%]\) & \(\mathrm{Uniform}(L=0.0,\ U=30.0)\) \\
        \hline
        \(F\) & Mass function nuisance parameter & \(\mathrm{Uniform}(L=1.0,\ U=7.5)\) \\
        \(s^2\) & Number density nuisance parameter \(\left[\mathrm{arcmin}^{-4}\right]\) & \(\mathrm{Uniform}(L=0.0,\ U=15.0)\) \\
        \(d\) & Heliocentric distance \([\mathrm{kpc}]\) & \(\mathrm{Gaussian}(\mu=d_{\mathrm{lit}},\ \sigma=\delta d_{\mathrm{lit}})\) \\
        \hline
    \end{tabular}
    \caption{List of all free parameters, their descriptions and the prior
             probability distributions used to bound their values.
             The first six are structural \limepy parameters
             (\Cref{sub:model_parameters}), while the next four define the mass
             function (\Cref{sub:mass_function_evolution}).
             The final three parameters aid in comparing models to observations
             (\Cref{sub:probability_distributions}).
             The prior distributions shown here, when not motivated by physical
             or model constraints (marked here by an asterisk; see
             \Cref{subsub:priors}), are chosen to bound a large enough area of
             parameter space containing all valid parameter values.
             The bounds here represent the largest extents used. In reality, the
             bounds may be reduced slightly during the sampling of certain
             clusters in order to improve the performance of the sampler, while
             still including a large area surrounding the most likely parameter
             values. The literature values and uncertainties used in the prior
             on the distance are taken from \citet{Baumgardt2021}.}
    \label{table:free_params}
    \end{table*}

    The overall structure of these models is controlled by the (dimensionless)
    central potential parameter \(\hat{\phi}_0\), which is used as a
    boundary condition for solving Poisson’s equation and defines how
    centrally concentrated the model is. The cluster model
    is spherical out to the truncation radius of the system, where its energy
    is reduced, mimicking the effects of the host galaxy's tides, which reduce
    the escape velocity of stars, making it easier for them to escape.
    The sharpness of this energy truncation is defined by the truncation
    parameter \(g\). Lower \(g\) values result in a more abrupt energy
    truncation, increasing up to models with the maximum possible finite extent
    at \(g=3.5\), while finite models with realistic values of \(\hat{\phi}_0\)
    are typically limited to \(g \lesssim 2.5\) \citep{Gieles2015}.

    The mass and size scales of the model can be expressed in any desired
    physical units by adopting corresponding values for the normalization
    constant \(A\) and the global velocity scale \(s\). We opt to scale the
    models to match observations using the parameters for total cluster mass
    \(M\) and 3D half-mass radius \(\rh\) as mass and size scales, which are
    used internally to compute the \(A\) and \(s\) scales.

    \limepy models allow for velocity anisotropy through an
    angular momentum term in the DF. With this term, the system is isotropic in
    the core, gains a degree of radial velocity anisotropy near the anisotropy
    radius \ra, and then becomes isotropic once more near the truncation radius.
    This parametrization mimics how GCs naturally develop
    radially-biased velocity anisotropy throughout their evolution
    as a result of two-body relaxation and tides \citep{Zocchi2016,Tiongco2016}.
    The two-body relaxation process drives the core of clusters to isotropy,
    however scattering (on preferentially radial orbits) of stars outside the
    core acts to increase the radial component of the velocity dispersion.
    Finally, a combination of the tidal torque from the host galaxy, which
    induces a transfer of angular momentum near the Jacobi radius to stellar
    orbits in the tangential direction \citep{Oh1992},
    and the preferential loss of stars on radial orbits \citep{Tiongco2016},
    act to increase the tangentiality of the outer stars,
    damping the amount of radial anisotropy and leading to a return to isotropy
    near the immediate edge of the system.
    The anisotropy radius \(\ra\) dictates the amount of radial
    velocity anisotropy present in the models.
    The smaller the value of \(\ra\), the more anisotropic the system. In the
    limit \(\ra \to \infty\), the models become entirely isotropic.
    In practice, models with \(\ra\) greater than the cluster truncation radius
    can be considered isotropic.

    The exact meaning of both the \(\hat{\phi}_0\) and \(\hat{r}_{\mathrm{a}}\)
    parameters depends on the definition of the mean mass \citep{Peuten2017}.
    In this work we adopt the global mean mass, that is, the mean mass of all
    stars in the entire cluster.

    The multimass version of the \limepy DF is defined by the sum of similar
    component DFs for each mass bin \(j\), with mass-dependent velocity
    (\(s_j\)) and anisotropy radius (\(\raj\)) scales. The
    mass-dependent velocity scaling captures the trend towards kinetic energy
    equipartition among stars of different masses and models the effects of mass
    segregation \citep{Gieles2015, Peuten2017, VHB2019}. The velocity scale is defined based on the parameter
    \(\delta\), such that \(s_j\propto s m_j^{-\delta}\), where \(s\) is defined
    as above. The mass-dependent anisotropy radius is defined in a similar
    fashion, using a parameter \(\eta\) (\(\raj\propto \ra m_j^{\eta}\)).
    For the analysis presented in this paper we have chosen to fix \(\eta\) to
    0, defining the anisotropy to be identical among all mass bins, the default
    assumption in multimass DF-based models.
    Our observations do not contain the information that would allow us to
    constrain the mass-dependence of the velocity anisotropy
    \citep[e.g.][]{Peuten2017}, and thus the \(\eta\) parameter.

    Finally, the constituent discrete mass components which approximate the
    mass spectrum of a GC are represented in the multimass
    \limepy models by the total (\(M_j\)) and mean (\(m_j\)) masses of each
    mass bin. These must be defined \textit{a priori} by external methods,
    based on the mass function (\(\alpha_1,\alpha_2,\alpha_3\)) and BH
    retention percentage (\(\BHret\)) parameters. The algorithm,
    which takes into account stellar evolution to predict the mean and total
    mass in stellar remnant bins, is described in detail in
    \Cref{sub:mass_function_evolution} below.

    External to the \limepy models themselves, we also employ a few extra
    parameters to aid in the fitting of the models to observations. These
    parameters are explained in more detail in
    \Cref{sub:probability_distributions}.

\subsection{Mass function evolution}\label{sub:mass_function_evolution}

    DF-based models, such as \limepy, compute the distribution of mass and
    velocity
    in a system in equilibrium. They are instantaneous ``snapshot'' models, and
    do not directly simulate any temporal astrophysical processes during their
    computation, including stellar evolution.
    As such, in order to determine the realistic mass populations for which the
    model will determine the phase-space distribution, we must incorporate a
    separate prescription for stellar evolution from an initial mass
    function, over the age of the cluster, to the present-day stellar and
    remnant mass functions.

    In keeping with the formulation of canonical IMFs
    \citep[e.g.][]{Kroupa2001}, we use a 3-component broken power law:
    \begin{equation}\label{eq:general_imf}
        \xi(m) \propto \begin{dcases}
            m^{-\alpha_1} & 0.1\ \Msun < m \leq 0.5\ \Msun, \\
            m^{-\alpha_2} & 0.5\ \Msun < m \leq 1\ \Msun, \\
            m^{-\alpha_3} & 1\ \Msun < m \leq 100\ \Msun, \\
        \end{dcases}
    \end{equation}
    where the \(\alpha_i\) parameters define the power-law `slope' of each
    component, and are allowed to vary freely during model fitting, and
    \(\xi(m) \dd{m}\) is the
    number of stars with masses within the interval \([m,\ m + \dd{m}]\).
    It should be noted here that our exact choices of break
    masses (0.1, 0.5, 1, 100 \Msun) are different than that of
    \citet{Kroupa2001}, to allow for a more specific study of the high-mass
    (\(m > \SI{1}{\Msun}\)) regime.

    To evolve the population of stars to the present day we follow the algorithm
    first described by \citet{Balbinot2018} and expanded upon in the
    \texttt{ssptools}\footnote{Available at
    \url{https://github.com/SMU-clusters/ssptools}}
    library. This method is summarized below.
    
    To begin, the rate of change of the number of main-sequence stars over time
    is given by the equation:
    \begin{equation}\label{eq:Ndot}
        \dot{N}(m_{\mathrm{to}}) = - \left.\dv{N}{m}\right|_{m_{\mathrm{to}}}
                          \left|\dv{m_{\mathrm{to}}}{t}\right|,
    \end{equation}
    where the amount of initial stars per unit mass (\(\dv*{N}{m}\)) at
    the turn-off mass (\(m_{\mathrm{to}}\)) is given by the IMF, and the rate of
    change of the turn-off mass can be derived by approximating the lifetime of
    main-sequence stars as a function of initial mass:
    \begin{equation}\label{eq:tms}
        t_{\mathrm{ms}} = a_0 \exp(a_1 m^{a_2}),
    \end{equation}
    where the \(a_i\) coefficients are interpolated from the Dartmouth
    Stellar Evolution Program models \citep{Dotter2007,Dotter2008}.
    This equation can then be inverted and differentiated to find the rate of
    change:
    \begin{equation}\label{eq:dmdt}
        \dv{m_{to}}{t} = \frac{1}{a_1 a_2} \frac{1}{t}
                             \left(\frac{\log(t/a_0)}{a_1}\right)^{1/a_2 - 1}.
    \end{equation}
    This set of equations dictates the amount of stars which evolve off the
    main sequence by the present-day cluster age \(t\). 

    In this work we use 30 discrete stellar mass bins, each
    logarithmically-spaced within the bounds of the three components defined by
    the IMF (5 bins in the low and intermediate-mass regime and 20 bins in the
    high-mass regime) and 30 identically spaced remnant mass bins, which
    are filled by the remnants resulting from stars evolving off the
    main-sequence.
    %
    As these stars evolve, the stellar remnants
    they will form (both in type and in mass), and thus the final remnant
    mass bins they will populate, depends on their initial mass
    and metallicity, and a functional initial-final mass relation (IFMR).

    The WD IFMR is computed as a 10th order polynomial:
    \begin{equation}
        m_{\mathrm{WD}} = \sum_{j=0}^{10} b_j m_i^j
    \end{equation}
    where \(m_i\) is the initial mass of the star, \(m_{\mathrm{WD}}\) is final
    mass of the formed remnant, and
    the coefficients \(b_j\) and the maximum initial mass which will
    form a WD are interpolated, based on
    metallicity, from the MIST 2018 isochrones \citep{Dotter2016,Choi2016}.

\begin{figure}
        \centering
        \includegraphics[width=\linewidth]{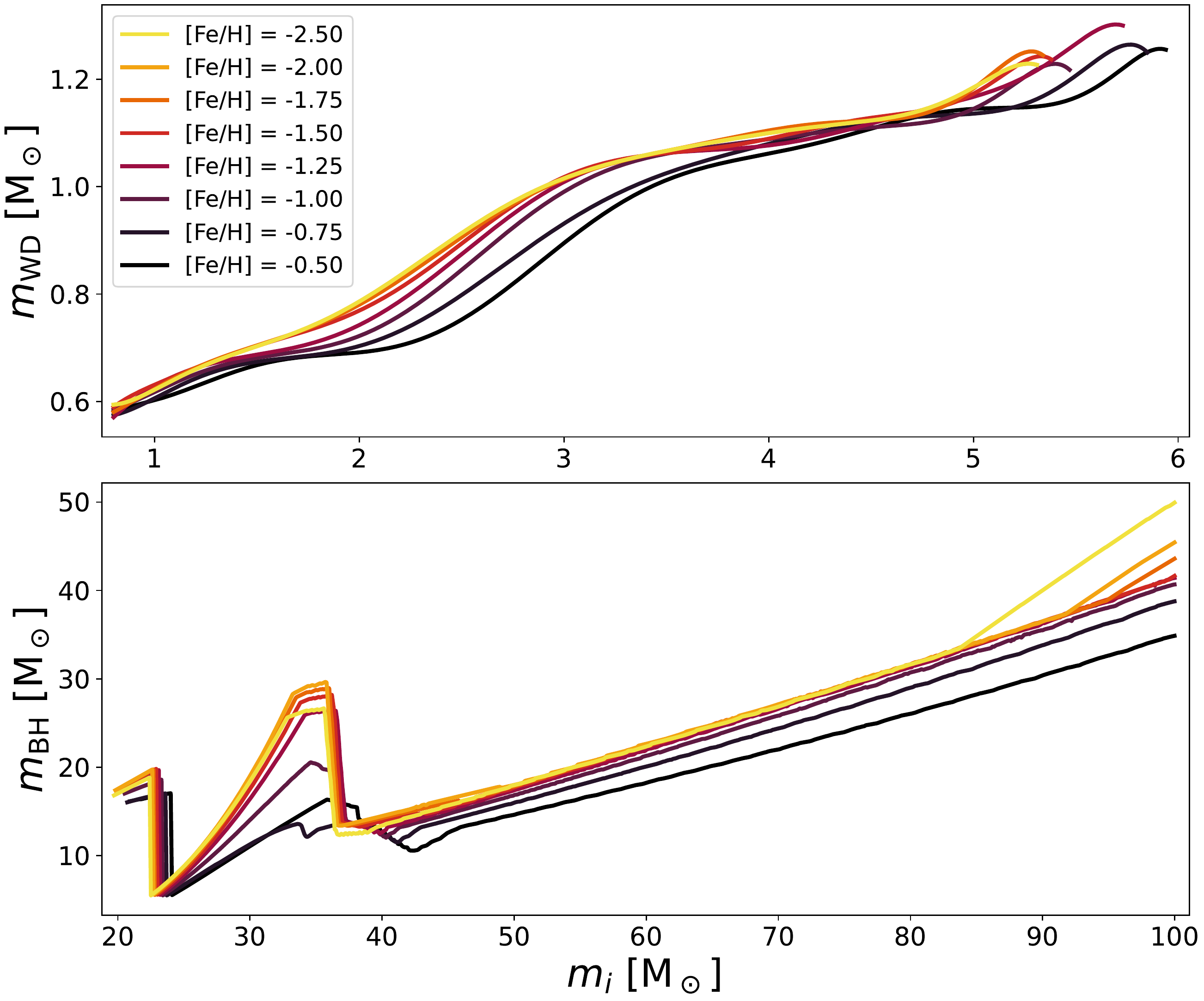}
        \caption{Adopted metallicity-dependent initial-final mass relations
                  for white dwarf (top panel) and black hole (bottom panel)
                  formation. Lower metallicities generally result in higher
                  final remnant masses.}
        \label{fig:ifmr}
    \end{figure}

    The BH IFMR, as well as the minimum initial
    mass required to form a BH, is interpolated directly from a grid of stellar 
    evolution library (SSE) models \citep{Banerjee2020}, using the rapid
    supernova scheme \citep{Fryer2012}, and is also dependent on metallicity.
    These relations are shown in Figure \ref{fig:ifmr}.
    All stars with initial masses between the WD and BH precursor
    masses are assumed to form neutron stars (NS). For simplicity, their final
    mass is always assumed to be \(1.4 \Msun\), regardless of the initial mass.

    The amount and final mass of these remnants (as dictated by Equation
    \ref{eq:Ndot}) must then be scaled downwards by an ``initial retention
    fraction'' \(f_{\mathrm{ret}}\), in order to mimic the loss of newly formed
    remnants due to natal kicks. For WDs we assume this is always equal to
    100\%. In this analysis, we assume a NS retention fraction of 10\%, as is
    common \citep[e.g.][]{Pfahl2002}, however, as shown in
    \citet{Henault-Brunet2020}, our results are insensitive to this exact value.

    The mass function evolution algorithm includes two more specific
    prescriptions for the loss of BHs, accounting for dynamical
    ejections in addition to natal kicks.

    Firstly the ejection of, primarily low-mass, BHs through natal kicks is
    simulated. We begin by assuming that the kick velocity is drawn from a
    Maxwellian distribution with a dispersion of \(\SI{265}{\km\per\s}\), as
    has been found for neutron stars \citep{Hobbs2005}.
    This velocity is then scaled down linearly by the ``fallback  fraction''
    \(f_{\mathrm{b}}\), the fraction of the precursor stellar
    envelope which falls back onto the BH after the initial supernova explosion.
    This fraction is interpolated from the same grid of SSE models used for the
    BH IFMR. The fraction of BHs retained in each mass bin is
    then found by integrating the Maxwellian kick velocity distribution
    from 0 to the system escape velocity. 
    The initial system escape velocity of each cluster was estimated by assuming 
    that about half of the initial cluster mass was lost through stellar
    evolution, while adiabatically expanding the cluster to a present-day
    half-mass radius a factor of two larger than the initial value,
    resulting in an initial escape velocity twice as large as the present-day
    value. A set of preliminary models were computed for all clusters, and the
    initial escape velocity was computed based on the best-fitting central
    density as \(v_{\mathrm{esc}}=2\sqrt{-2\phi_0}\), where \(\phi_0\) is the
    central potential.
    It should be noted that clusters with an escape velocity
    \(\gtrsim \SI{100}{\km\per\s}\) will retain nearly all BHs
    \citep*{Antonini2019}.

    Black holes are also ejected over time from the core of GCs due to dynamical
    interactions with one another \citep[e.g.][]{Breen2013a,Breen2013b}. This
    process is simulated through the removal of BHs, beginning with the heaviest
    mass bins (with larger gravitational interaction cross-sections)
    through to the lightest \citep{Morscher2015,Antonini2020}, until the
    combination of mass in BHs lost through both the natal
    kicks and these dynamical ejections leads to a retained mass in BHs
    corresponding to the percentage of the initial mass in BHs specified
    by the BH mass retention fraction parameter
    (\(\mathrm{BH}_{\mathrm{ret}}\)).

    The final avenue for cluster mass loss is through the escape of stars and
    remnants driven by two-body relaxation and lost to the host
    galaxy. Such losses, in a mass segregated cluster, are dominated by
    the escape of low-mass objects from the outer regions of the cluster.
    Determining the overall losses through this process is a complicated task,
    dependent on the dynamical history and orbital evolution of the cluster,
    which we do not attempt to model here.
    We thus opt to ignore this preferential
    loss of low-mass stars and do not further model the escape of any
    stars, apart from through the processes described above.
    This means that the low-mass \(\alpha\) exponents determined here may,
    in most cases, describe most accurately the PDMF rather than the low-mass
    IMF of our clusters. This is discussed in more detail in
    \Cref{ch:mass_functions}.


\section{Cluster data}\label{sec:cluster_data}

    In this work, we determine best-fitting model parameters for 37 Milky
    Way globular clusters through the comparison of the phase-space
    distribution of stars in the \limepy models to analogous observations of
    GC structure and kinematics.

\subsection{Cluster selection}

    The clusters analyzed in this work were selected from the population of
    Milky Way GCs in order to best study the possible relationship of the
    mass function with metallicity. To do so, we choose clusters over a
    range of metallicities (taken from \citealp{Harris}), with most clusters
    in our sample being metal-poor
    (\([\chem{Fe}/\chem{H}] \lesssim -1.0 \)). The greatest
    discerning factor used in cluster selection was the quantity and quality of
    data available.
    We searched the catalogue of observational datasets presented by 
    \citet{Baumgardt2017,Baumgardt2018}
    and \citet{Baumgardt2022}\footnote{Available at
    \url{https://people.smp.uq.edu.au/HolgerBaumgardt/globular/}}
    for clusters with a combination of adequate
    mass function depth and radial coverage from HST photometry, and sufficient
    kinematic data to constrain the models.
    These selection criteria lead to the choice of 37 final clusters.

\subsection{Datasets}\label{sec:datasets}

    Models are fit to all chosen GCs through comparison with a variety of
    observational datasets, which help directly constrain both the 
    distribution of visible cluster stars through direct stellar number counts,
    and the overall total mass of the cluster through accurate kinematic
    profiles. This, in turn, provides indirect
    constraints on the amount and distribution of dark mass (in both faint
    low-mass stars and dark remnants) making up the difference between
    the visible and total mass, as, together with mass segregation, the
    possible distribution of cluster mass among different components
    has limited flexibility. Key model parameters, in particular
    \(\alpha_3\), which sets the amount of high-mass stars and remnants in the
    models, can thus be constrained by this combination of datasets.
    We utilize a large number of observables from various sources, while aiming
    to provide as much homogeneity between clusters as possible.
    All literature sources used for each cluster are listed in
    \Cref{appendix:data_sources}.

\subsubsection{Proper motions}

    Radial profiles of the dispersion of proper motions (PMs) of cluster stars
    are used to constrain the cluster velocity dispersion profiles, and in turn
    the total cluster mass and its distribution.
    By incorporating the kinematics in both the radial and tangential
    directions in the plane of the sky, we are also able to constrain the
    amount of velocity anisotropy in the system. We define these components,
    on the sky, such that the radial component is positive outwards from the
    cluster centre, and the tangential component is positive in the
    counterclockwise rotational direction on the sky. Given the proper motions
    of a star in a cluster-centred orthographic projection (e.g. equation 2 in
    \citealt{GaiaCollaboration2018b}), the radial (\(\mu_R\)) and tangential
    (\(\mu_T\)) components are defined as:
    \begin{align}
        \mu_R \equiv \frac{(x \mu_x + y \mu_y )}{R},
        && \mu_T \equiv \frac{(y \mu_x - x \mu_y )}{R},
    \end{align}
    where \(x\), \(y\), \(\mu_x\) and \(\mu_y\) are the orthographic positions
    and proper motions and \(R=\sqrt{x^2+y^2}\) is the projected distance from
    the cluster centre, which is taken from \citet{Baumgardt2017}.


    We extract our own PM dispersion profiles in both components from Gaia Early
    Data Release 3 (EDR3; \citealp{GaiaCollaboration2021}) proper motions for
    all clusters.\footnote{Extracted Gaia EDR3 PM dispersion profiles for all
    clusters are available for download from
    \url{https://github.com/nmdickson/GCfit-results}}.
    The catalogue of cluster stars, along with their membership probabilities,
    is taken from \citet{Vasiliev2021}. 
    Following the conclusions of \citet{Vasiliev2021}, in order to account for
    underestimations in the statistical uncertainty of proper motions of Gaia
    sources in dense regions, we scale the PM uncertainties of each star by a
    density-dependent factor \(\eta\):
    \begin{equation}
        \eta = \left(1 + \frac{\Sigma}{\Sigma_0}\right)^\zeta,
    \end{equation}
    where \(\Sigma\) is the nearby stellar density,
    \(\Sigma_0 = \SI{10}{stars\per arcmin^2}\) and \(\zeta=0.04\)
    (from Table 1 in \citealp{Vasiliev2021}).
    We then follow a similar methodology to \citet{Vasiliev2019c} and
    \citet{Vasiliev2021} to construct radially binned dispersion profiles in
    both directional components by fitting a multivariate Gaussian distribution
    to the proper motions of all the stars in each bin which pass the quality
    flags described in section 2 of \citet{Vasiliev2021}.


    We supplement the Gaia proper motion datasets of specific clusters, where
    further PM studies are available from the Hubble Space Telescope (HST).
    %
    \citet{Libralato2022} presented profiles of proper motion dispersions in the
    central regions of 57 globular clusters, based on archival HST photometry.
    This catalogue overlaps with our sample for 35 clusters, 
    in which case we utilize both the radial \(\sigma_R\) and tangential
    \(\sigma_T\) components.
    %
    For two of the clusters in our sample not covered by \citet{Libralato2022}
    (\NGC5139 and \NGC6266) we instead utilize the the total
    dispersion \(\sigma = \sqrt{(\sigma_T^2 + \sigma_R^2) / 2}\) and
    anisotropy ratio \(\sigma_T / \sigma_R\) profiles presented by
    \citet{Watkins2015}, based on the HST catalogues of \citet{Bellini2014}.
    %
    The coverage of the core of \NGC6723 is also extended
    by the dispersion profiles of \citet{Taheri2022} (using Gemini South GeMS).

\subsubsection{Line-of-sight velocities}

    The kinematic data is also supplemented by line-of-sight (LOS) velocity
    dispersion profiles, providing a 3-dimensional view of the
    cluster dynamics.

    The majority of the LOS dispersion profiles used come from compilations of
    different surveys and programs. \citet{Baumgardt2017} gathered, from the
    literature, 95 publications with large enough LOS velocity datasets for 45
    GCs, and \citet{Baumgardt2018} expanded on this catalogue by including
    additional ESO/Keck archival data of the LOS velocities of
    stars in 90 GCs. In both cases, the different datasets were combined by
    shifting them to the cluster's mean radial velocity.
    \citet{Baumgardt2019a} derived the velocity dispersion profiles of 127 GCs
    using the Gaia DR2 radial velocity data. This catalogue of stars was matched
    to that of \citet{Baumgardt2018}, and scaled to a common mean velocity.
    Finally, this catalogue was enhanced again by \citet{Baumgardt2022}
    with the inclusion of data from various more recent large scale
    radial-velocity surveys.
    %
    \citet{Dalgleish2020} supplemented this work with radial velocity
    measurements in 59 GCs from the WAGGS survey, using the WiFeS integral field
    spectrograph.
    These datasets were further complemented in the cores of 22 clusters by
    the LOS dispersion profiles presented by \citet{Kamann2018}, who gathered
    data within the half-light radius of 22 GCs using the MUSE
    integral-field-unit spectrograph on the VLT.
    Further coverage of the central region of \NGC{6266} is provided by the
    profiles presented by \citet{Lutzgendorf2013},
    based on observations by the VLT/FLAMES integral-field-unit spectrograph,
    and of \NGC{6362} by \citet{Dalessandro2021}, based on VLT/MUSE
    observations.

\subsubsection{Number density profiles}\label{sub:number_density_profiles}

    Radial profiles of the projected number density of stars in our GCs
    are vital in constraining the spatial structure and concentration of the
    clusters.

    The projected number density profiles of all clusters are taken from
    \citet{deBoer2019}, who utilized counts of member stars from Gaia DR2,
    binned radially, for 81 Milky Way clusters. Membership was determined, for
    stars up to a faint magnitude limit of \(G = 20\), based on the Gaia
    proper motions.
    To aid with the coverage of the cluster centres, where Gaia is incomplete
    and struggles with crowding in all but the least dense GCs, the authors
    stitched the Gaia profiles together with profiles from HST photometry
    \citep{Miocchi2013} and a collection of ground-based surface brightness
    profiles \citep{Trager1995}. These profiles from the literature were scaled
    to match the Gaia profiles in the regions where they overlap, with the
    final profile being constructed of Gaia counts in all regions with a
    density lower than \(10^5\ \mathrm{stars}/\mathrm{deg}^2\) and literature
    profiles otherwise.
    \citet{deBoer2019} also computed a constant background contamination level
    for each cluster, computed as the average stellar density between 1.5 and 2
    Jacobi radii, which we subtract from the entire profile before fitting.

\subsubsection{Mass functions}

    To provide constraints on the global present-day mass function of
    the clusters, the degree of mass segregation and the total mass in visible
    stars, we compare our models against measurements of the stellar mass
    function in radial annuli and mass bins obtained from deep HST photometry.

    The mass function data for each cluster was derived from archival HST
    photometry by \citet{Baumgardt2022} and includes data from large-scale
    archival surveys \citep[e.g.][]{Sarajedini2007,Simioni2018}.
    Stellar photometry and completeness correction of the data was done using
    \texttt{DOLPHOT} \citep{Dolphin2000,Dolphin2016}. Stellar number counts
    were then derived as a function of stellar magnitude and distance from the
    cluster centre and were then converted into stellar mass functions through
    fits to DSEP isochrones \citep{Dotter2008}. See \citet{Baumgardt2022} for
    more details on the extraction and conversion of these mass functions
    \footnote{To fit the isochrones, \citet{Baumgardt2022} begins with the
    cluster heliocentric distances from \citet{Baumgardt2021}, and allows
    the distance to vary slightly. This is similar to our methodology (see
    \Cref{subsub:priors}), but may result in slightly different final
    distances to ours. This may introduce a slight inconsistency
    but, given the distances are all in agreement within uncertainties, will
    have a negligible impact on our results.}.
    The compilation of images is made up of several HST fields
    for each cluster, at varying distances from the cluster centres. The
    observations typically cover stars within a mass range of
    \(\sim\SI{0.16}-\SI{0.8}{\Msun}\). The large radial and mass ranges covered
    allow us to constrain the varying local stellar mass function as a function
    of distance from the cluster centre, and therefore the degree of mass
    segregation in the cluster.


\section{Model Fitting}\label{ch:model_fitting}

    The models described in \Cref{sec:models} are constrained by the data
    described in \Cref{sec:cluster_data} in order to provide
    distributions of the best-fitting model parameters that describe each
    cluster, which are determined through Bayesian parameter estimation
    techniques.

\subsection{Probability distributions}\label{sub:probability_distributions}

    Given a model \(M\), the probability associated with a given set of model
    parameters \(\Theta\), subject to some observed data \(\mathcal{D}\) is
    given by the Bayesian posterior:
    \begin{equation}\label{eq:bayesian_post}
        P(\Theta \mid \mathcal{D}, M) = \frac{P(\mathcal{D} \mid \Theta,M)
                                        P(\Theta \mid M)}{P(\mathcal{D} \mid M)}
                         = \frac{\mathcal{L}(\Theta) \pi(\Theta)}{\mathcal{Z}},
    \end{equation}
    where \(\mathcal{L}\) is the likelihood, \(\pi\) is the prior and
    \(\mathcal{Z}\) is the evidence.

\subsubsection{Likelihood}\label{subsub:likelihood}

    In this work, the total log-likelihood function \(\ln(\mathcal{L})\), for
    all data \(\mathcal{D}\) considered for a certain cluster, is
    given simply by the summation of all log-likelihood functions for each
    individual dataset \(\mathcal{D}_i\):
    \begin{equation}
        \ln(\mathcal{L}) = \sum_i^{\mathrm{datasets}}\ln(P(\mathcal{D}_i\mid\Theta))
                         = \sum_i \ln(\mathcal{L}_i(\Theta))),
    \end{equation}
    and each observational dataset, as described in \Cref{sec:datasets},
    has its own component likelihood function \(\ln(\mathcal{L}_i)\), detailed
    below.

    In order to compare all observed quantities with model predictions,
    certain quantities which involve angular units (radial distances,
    proper motions, cluster radii, etc.) must be converted to the projected,
    linear model lengths. 
    To do so, we introduce the heliocentric distance to the GC as a new free
    parameter \(d\), and use the velocity and position conversions:
    \begin{equation}
        v_T = 4.74\, \mathrm{km/s}\ \frac{d}{\mathrm{kpc}}
                                  \ \frac{\mu}{\mathrm{mas/yr}},
    \end{equation}
    \begin{equation}
        r = 2\ d  \tan\left(\frac{\theta}{2}\right),
    \end{equation}
    where \(v_T\) is the plane-of-the-sky velocity, \(\mu\) is the observed
    proper motion, \(r\) is the distance to the cluster centre in projection and
    \(\theta\) is the observed angular separation.
    
    In all likelihood functions below, the modelled quantities, unless
    otherwise stated, are taken from the mass bin most closely corresponding to
    the masses of stars observed in each dataset.

    \paragraph{Velocity dispersion profiles}

    The likelihood function used for all velocity dispersions (LOS and PM) is a
    Gaussian, over a number of dispersion measurements at different projected
    radial distances:
    \begin{equation}
        \ln(\mathcal{L}_i) = \frac{1}{2} \sum_j
            \left[
                \frac{(\sigma_{j,\mathrm{obs}}
                - \sigma_{j,\mathrm{model}})^2}
                {\delta\sigma_{j,\mathrm{obs}}^2}
                - \ln(\delta\sigma_{j,\mathrm{obs}}^2)
            \right],
    \end{equation}
    where \(\sigma_j \equiv \sigma(r_j)\) corresponds to the dispersion at a
    distance \(r_j\) from the cluster centre, with corresponding uncertainties
    \(\delta\sigma_j \equiv \delta\sigma(r_j)\). Dispersions with subscript
    \textit{obs} correspond to the observed dispersions and uncertainties, while
    subscript \textit{model} corresponds to the predicted model dispersions.

    \paragraph{Number density profiles}

    The likelihood function used for the number density profile datasets is a
    modified Gaussian likelihood.

    The translation between the surface brightness measurements and discrete
    star counts (both considered for the number density profiles, as discussed
    in \Cref{sub:number_density_profiles}), is difficult to quantify exactly.
    To compare
    star counts above a magnitude limit to the integrated light of a surface-brightness profile would require precise knowledge of the mass-to-light
    ratio for each mass bin, which is an uncertain quantity, especially for
    evolved stars. To account for this in the fitting procedure, the model is
    actually fit on the \textit{shape} of the number density profile, rather
    than on the absolute star counts. To accomplish this the number density
    profile of the model is scaled to have the same mean value as the observed
    profiles. As in \citet{Henault-Brunet2020}, the constant
    scaling factor \(K\) is chosen to minimize the chi-squared:
    \begin{equation}
        K = \frac{\sum\limits_j \Sigma_{j,\mathrm{obs}} \Sigma_{j,\mathrm{model}}
                  / \delta\Sigma^2_j}
                 {\sum\limits_j \Sigma_{j,\mathrm{model}}^2 / \delta\Sigma^2_j},
    \end{equation}
    where \(\Sigma_j \equiv \Sigma(r_j)\) are the modelled and observed
    number density, with respective subscripts, at a distance \(r_j\) from the
    cluster centre.

    We also introduce an extra ``nuisance''
    parameter (\(s^2\)) to the fitting. This parameter is added in quadrature,
    as a constant error over the entire profile, to the observational
    uncertainties to give the overall error \(\delta\Sigma\):
    \begin{equation}
        \delta\Sigma^2_j = \delta\Sigma_{j,\mathrm{obs}}^2 + s^2.
    \end{equation}
    This parameter adds a constant uncertainty component over the entire radial
    extent of the number density profile, effectively allowing for small
    deviations in the observed profiles near the outskirts of the cluster.
    This enables us to account for certain processes not captured by our models,
    such as the effects of potential escapers \citep{Claydon2017,Claydon2019}.

    The likelihood is then given in similar fashion to the dispersion profiles:
    \begin{equation}
        \ln(\mathcal{L}_i) = \frac{1}{2} \sum_j
            \left[
                \frac{(\Sigma_{j,\mathrm{obs}}
                - K\Sigma_{j,\mathrm{model}})^2}{\delta\Sigma^2_j}
                - \ln(\delta\Sigma^2_j)
            \right].
    \end{equation}

    \paragraph{Mass functions}

    To compare the models against the mass function datasets,
    the local stellar mass functions are extracted from the models within
    specific areas in order to match the observed MF data at different projected
    radial distances from the cluster centre within their respective HST fields.

    To compute the stellar mass functions, the model surface density in a given
    mass bin \(\Sigma_k(r)\) is
    integrated, using a Monte Carlo method, over the area \(A_j\), which covers
    a radial slice of the corresponding HST field from the projected distances
    \(r_j\) to \(r_{j+1}\). This gives the count \(N_{\mathrm{model},k,j}\) of
    stars within this footprint \(j\) in the mass bin \(k\):
    \begin{equation}
        N_{\mathrm{model}, k, j} = \int_{A_j} \Sigma_k(r) dA_j.
    \end{equation}
    This star count can then be used to compute the Gaussian likelihood:
    \begin{equation}
        \ln(\mathcal{L}_i) = \frac{1}{2}
            \sum_j^{\substack{\mathrm{radial}\\\mathrm{bins}}}
            \sum_k^{\substack{\mathrm{mass}\\\mathrm{bins}}}
            \left[
                \frac{(N_{\mathrm{obs},k,j} - N_{\mathrm{model},k,j})^2}
                        {\delta N_{k,j}^2}
                  - \ln(\delta N_{k,j}^2)
            \right],
    \end{equation}
    which is computed separately for each HST program considered.

    The error term \(\delta N_{k,j}\) must also account for unknown and
    unaccounted for sources of error in the mass function counts, as well as the
    fact that our assumed parametrization of the global mass function
    (\cref{eq:general_imf}) may not be a perfect representation of the data.
    Therefore we include another nuisance parameter (\(F\)) which scales up
    the uncertainties:
    \begin{equation}
        \delta N_{k,j} = F \cdot \delta N_{\mathrm{obs},k,j}.
    \end{equation}
    This scaling, rather than adding in quadrature as with the \(s^2\)
    nuisance parameter, boosts the errors by a constant factor.
    This allows it to capture additional unaccounted-for uncertainties
    (e.g. in the completeness correction or limitations due to the simple
    parametrization of the mass function) across the full range of values of
    star counts, while simply adding the same error in quadrature to all values
    of star counts would lead to negligible error inflation in regions with
    higher counts.

\subsubsection{Priors}\label{subsub:priors}

    The prior probability distribution \(\pi\) for our set of model parameters
    \(\Theta\) is given by the product of individual, independent priors for
    each parameter in \(\Theta\):
    \begin{equation}
        \pi(\Theta) = \prod_i^{N_{\mathrm{params}}} \pi_i (\theta_i).
    \end{equation}
    The priors for individual parameters can take a few possible forms.

    \textit{Uniform}, or flat, priors are used to provide an
    uninformative prior to most parameters. The uniform distribution is defined
    as constant between two bounds $(L, U)$, with a total probability normalized
    to unity:
    \begin{equation}
        \pi_i (\theta_i) =
        \begin{cases}
            \displaystyle({U-L})^{-1} & \text{for}\, L \le \theta_i \le U, \\
            0 & {\text{otherwise}}.
        \end{cases}
    \end{equation}

    The upper and lower bounds are chosen, for most parameters, to simply
    bound a large enough area of parameter space containing all valid
    parameter values, whereas for certain parameters the bounds are specifically
    set to disallow values outside a certain range. All parameters except the
    heliocentric distance (described below) use uniform priors.

    The truncation parameter \(g\) is limited to values between 0 and 3.5 for
    all clusters, corresponding to the absolute limit of models of finite
    extent \citep{Gieles2015}.

    The mass-dependant velocity scale \(\delta\) is given an upper limit
    of 0.5, corresponding to the typical value reached by fully
    mass segregated cluster, and a lower limit of 0.3.
    Comparisons between \limepy models and \Nbody simulations of GCs have
    shown that even less evolved clusters, still containing a large number
    of BHs, are best-fit by a \(\delta \sim 0.35\) \citep{Peuten2017}.
    In this work, not even \omegacen reaches the low limit of our prior
    range.

    Finally, the mass function exponents \(\alpha_i\) are limited to reasonable
    regimes. The low and intermediate mass components \(\alpha_1\) and
    \(\alpha_2\)
    are given bounds between -1 and 2.35, confining the MF to remain shallower
    than the canonical high-mass IMF, and allowing for an increasing mass
    function with increasing masses, which may best describe the most evolved
    clusters. The high-mass exponent \(\alpha_3\) is restricted to values
    between 1.6 and 4.0. The lower-bound of 1.6 is chosen as it has been shown
    that clusters this ``top-heavy'' are expected to have dissolved by the
    present day \citep{weatherford2021,Haghi2020}.
    The upper limit of 4 is chosen as,
    above this value, lower-mass globular clusters will contain very few heavy
    remnants and no neutron stars or black holes, in contradiction with
    observations of stellar remnants within clusters.
    All exponents are also required to decrease from the lower to the higher
    mass regimes, such that \(\alpha_1 \leq \alpha_2 \leq \alpha_3\),
    following currently observed constraints, Although note that tests with
    this final rule relaxed resulted in no significant differences.

    {Gaussian} priors are used for the parameters which are informed by
    previous and independent analyses, and take the form of a Gaussian
    distribution centred on the reported value \(\mu\) with a width of
    corresponding to the reported uncertainty \(\sigma\):
    \begin{equation}
        \pi_i (\theta_i)  = \frac{1}{\sigma \sqrt{2\pi}}
        e^{-\frac{1}{2} \left(\frac{\theta_i-\mu}{\sigma}\right)^{2}}.
    \end{equation}

    In particular for this analysis, we adopt a Gaussian prior for the distance
    parameter \(d\), with a mean and standard deviation taken from
    \citet{Baumgardt2021}. This allows the distance to vary in order to
    accommodate other observational constraints used in this work, while
    still being strongly influenced by the robust value obtained through the
    averaging of a variety of distance determinations from different methods
    from by \citet{Baumgardt2021}.

    The priors used for all parameters are listed in \Cref{table:free_params}.

\subsection{Sampling}\label{sec:sampling}

    The posterior probability distribution \(P(\Theta \mid \mathcal{D}, M)\)
    of the parameter set \(\Theta\) cannot be solved analytically, but must be
    estimated through numerical sampling techniques, which aim to generate a
    set of samples that can be used to approximate the posterior distribution.

    Nested sampling \citep{Skilling2004,Skilling2006} is a Monte Carlo
    integration method, first proposed for estimating the Bayesian
    evidence integral \(\mathcal{Z}\), which works by iteratively integrating
    the posterior over the shells of prior volume contained within nested,
    increasing iso-likelihood contours.

    Samples are proposed randomly at each step, subject to a minimum likelihood
    constraint corresponding to the current likelihood contour. This sampling
    proceeds from the outer (low-likelihood) parameter space inwards, until
    the estimated remaining proportion of the evidence
    integral, which arises naturally from the sampling, reaches a
    desiredly small percentage. This well-defined
    stopping criterion is a great advantage of nested sampling, as in most other
    sampling methods convergence can be difficult to ascertain.


    Nested sampling has the benefit of flexibility, as the independently
    generated samples are able to probe complex posterior shapes, with little
    danger of falling into local minima, or of missing distant modes.
    It also does not depend, like many other sampling methods, on a
    choice of initial sampler positions, and will always cover the entire
    prior volume.
    In cases of well-defined priors and smoothly transitioning posteriors, as
    is the case in this work, the sampling efficiency can
    exceed that of the typical Markov chain Monte Carlo (MCMC) samplers.


    Dynamic nested sampling is an extension of the typical nested
    algorithm designed to re-tune the sampling to more efficiently estimate the
    posterior \citep{Higson2019}. This algorithm effectively functions by
    spending less time probing the `outer' sections of the prior volume which
    have little impact on the posterior.
    In this work, we have chosen to utilize dynamic nested sampling for
    its speed and efficiency, and to ensure that no separate, distant modes
    in the posterior are missed.


    All methodology in this work, from data collection to model fitting, is
    handled by the software library and fitting pipeline \texttt{GCfit}
    \footnote{Available at \url{https://github.com/nmdickson/GCfit}},
    which was created to facilitate the fitting of \limepy models to a number
    of observables through a parallelized sampling procedure.
    All nested sampling is handled by the \texttt{dynesty} software package
    \citep{Speagle2020}. The sampler is run, for all clusters, using the default
    (multi-ellipsoid bounded, random-walk) dynamic sampling (see
    \citealt{Speagle2020} for more details). The sampling is continued until
    it reaches an effective sample size (ESS; \citealp{kish1965}) of at least
    5000:
    \begin{equation}
        \mathrm{ESS} = \frac{\left(\sum_{i=1}^n w_i\right)^2}{\sum_{i=1}^n w_i^2},
    \end{equation}
    where \(w_i\) is the importance weight of the sample \(i\) in the set of
    generated samples.



\section{Results}\label{ch:results}

We present in this section the results of the fits based on the methodology
of \Cref{ch:model_fitting}. First we introduce the resulting posterior
probability distributions of all model parameters, and the corresponding fits
they give to the relevant data. We then briefly discuss the distribution
between clusters of some structural parameters of interest.
The stellar mass functions of the clusters are explored in more detail in
\Cref{ch:mass_functions}.

\subsection{Fitting Results}\label{sec:fitting_results}

\subsubsection{Parameter distributions}
    

    The set of weighted samples retrieved from the nested sampler, after
    sampling until the stopping condition described in \Cref{sec:sampling}, are
    used to construct posterior probability distributions for all model
    parameters.


    \begin{figure*}
        \centering
        \includegraphics[width=\linewidth]{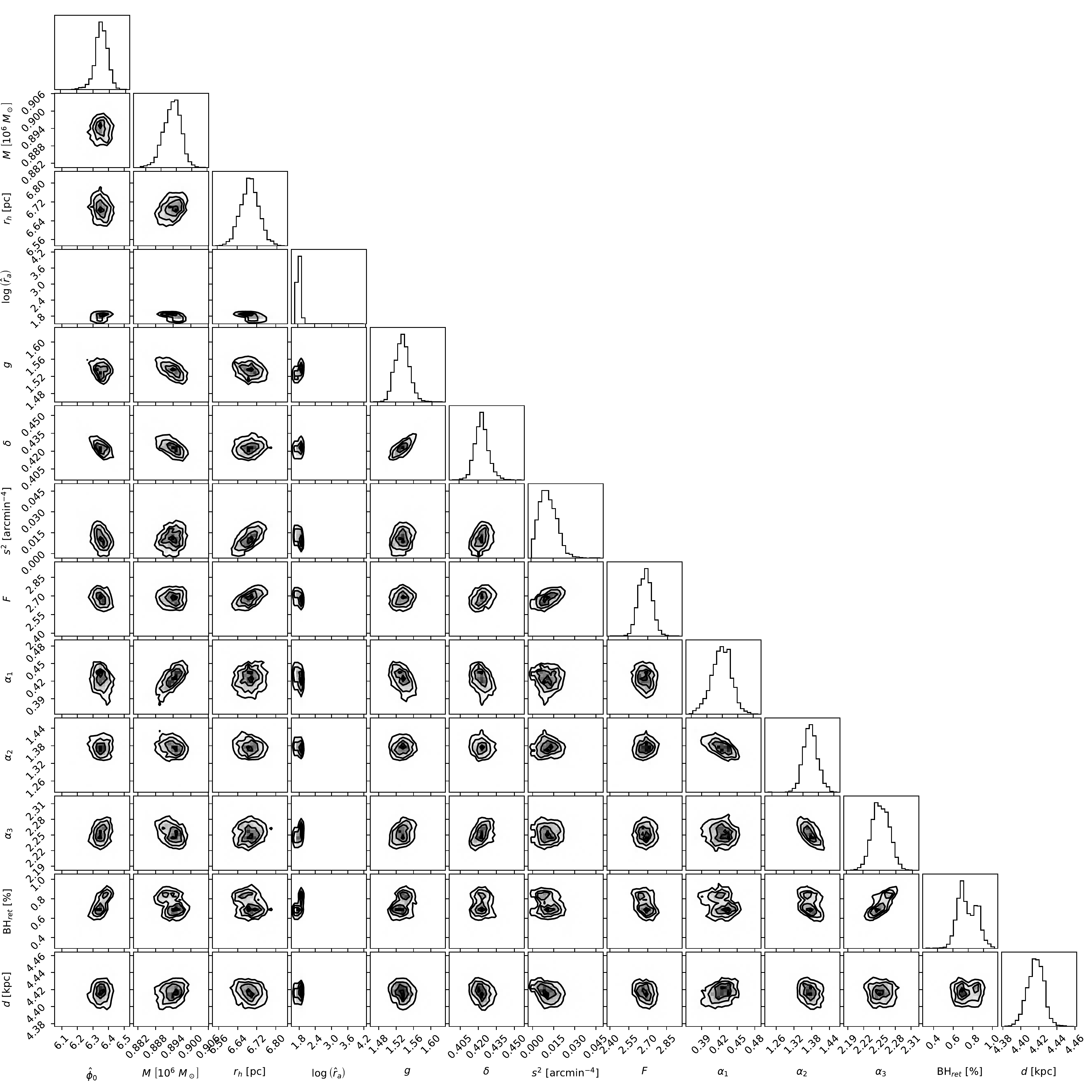}
        \caption{Marginalized and 2D projections of the posterior probability
                 distributions of all model parameters for the fit to
                 \NGC104. Contours indicate \(1\sigma\), \(2\sigma\) and
                 \(3\sigma\) levels on the 2D posterior probability
                 distributions.}
        \label{fig:NGC0104_marginals}
    \end{figure*}

    \Cref{fig:NGC0104_marginals} shows an example of the resulting
    posterior distributions for the cluster \NGC104. 
    The best-fitting parameter values for all clusters can be found in
    \Cref{table:best_fitting_params}\footnote{All fit results, figures and
    tables from this paper are also available for all clusters by download from
    \url{https://github.com/nmdickson/GCfit-results}.
    Figures showing the fits for all clusters in the sample are also
    available as supplementary material in the electronic version}.


    The vast majority of marginalized posterior distributions for the cluster
    parameters follow a unimodal and approximately Gaussian distribution.
    The marginalized posterior probability distribution of some parameters are
    skewed towards or hitting the boundaries of the prior ranges, however, as
    indicated in \Cref{subsub:priors}, this is only allowed to occur for
    parameters with physically motivated prior boundaries.
    
    The posterior parameter distributions of one cluster (\NGC6723) are not
    single Gaussians, but instead show two separate peaks, both containing
    comparable posterior probability. This cluster will be
    discussed in more detail in \paperII, as these models differ most
    significantly in their BH populations. In all figures in
    this paper this cluster may appear as a single point with a very
    large errorbars in one direction, due to the fact that one of these peaks
    is larger than the other, and the median of the
    entire distribution falls entirely within this peak.

    Two parameters (\(\log(\hat{r}_{\mathrm{a}})\) and \(\BHret\)) often have a
    broader posterior probability distribution. The anisotropy radius may be
    unconstrained above a certain minimum value, illustrating the fact that all
    values of the anisotropy radius greater than the truncation radius
    effectively lead to an entirely isotropic cluster. The BH
    retention fraction may be completely unconstrained in models with a very
    small number of BHs initially formed  (e.g. due to a "top-light"
    mass function), in which case the fraction of BHs retained has a
    negligible effect on the models. These parameters are examined below and in
    \paperII, respectively.

    A fraction of our clusters are core-collapsed, and they are expected to not
    retain any significant populations of BHs \citep{Giersz2009, Breen2013a}.
    However, our best-fitting models of four such clusters
    (\NGC6266, \NGC6624, \NGC6752, \NGC7078) do possess BHs, and
    may not be physical. Core-collapsed clusters have a cusp in the inner
    surface brightness profile which is difficult to reproduce with the \limepy
    models, which are cored. We therefore do not trust the result for BH
    retention of these clusters as we suspect that \(\BHret\) was used as
    an additional degree of freedom in an attempt to describe the inner
    profiles. In these cases we recompute the models, this time
    with the amount of retained BHs at the present
    day fixed to 0 (by fixing the \(\BHret\) parameter to 0\%).
    These models are used, for these four clusters, in all analysis presented
    in this paper.
    This phenomenon, these models and the limitations of our \limepy models in
    representing core-collapsed clusters will all be examined and discussed in more
    detail in \paperII. However, both sets of models demonstrate good fits
    to the data, and there was no significant change in the best-fit mass
    function slopes or in any of the correlations presented in this paper, when
    considering either set of models.

\subsubsection{Best-fitting models}

\begin{figure*}
        \centering
        \includegraphics[width=\linewidth]{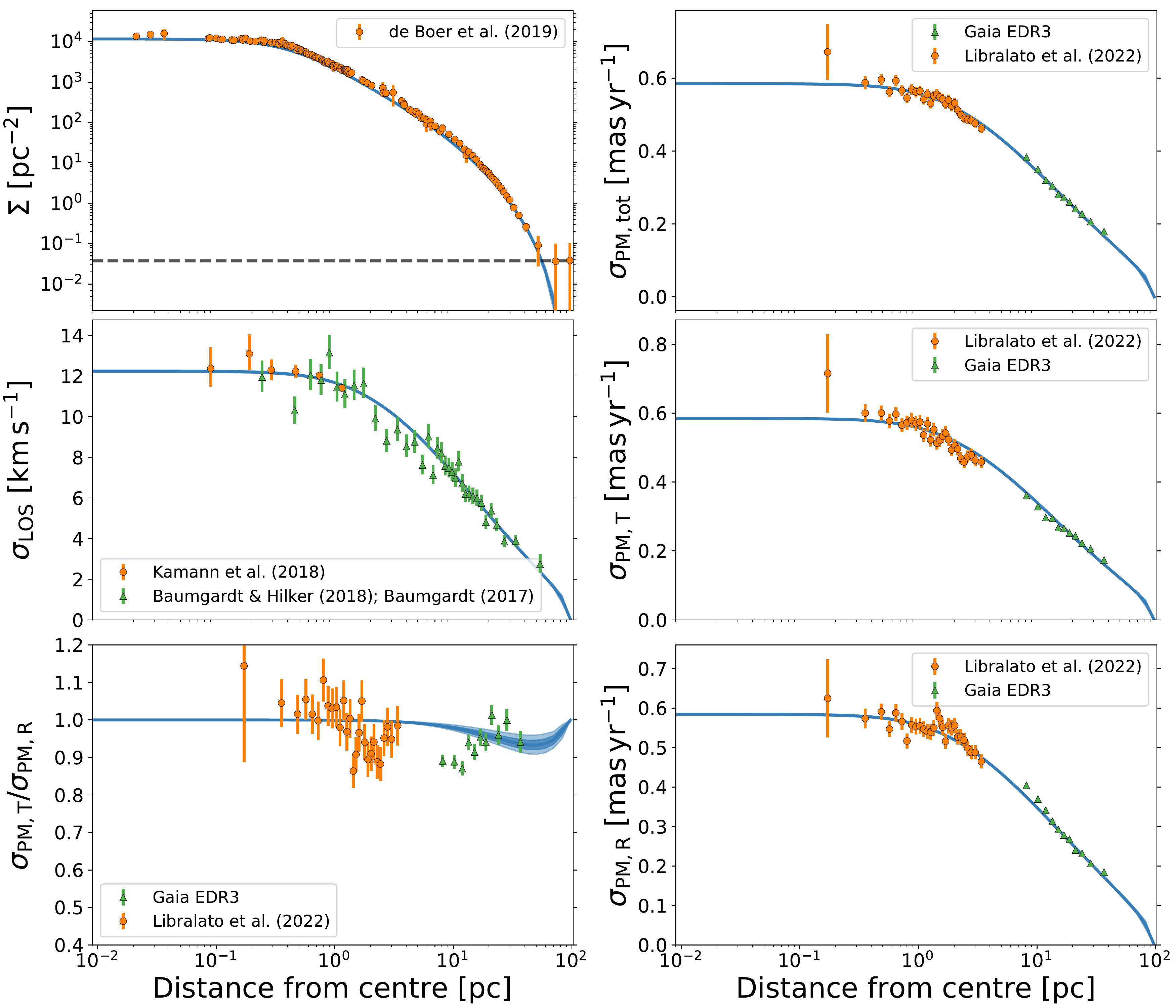}
        \caption{Model radial profiles (blue contours) of
                 surface number density (\(\Sigma\)),
                 line-of-sight velocity dispersions (\(\sigma_{\mathrm{LOS}}\)),
                 total (\(\sigma_{\mathrm{PM},\mathrm{tot}}\)),
                 radial (\(\sigma_{\mathrm{PM},\mathrm{R}}\)) and
                 tangential (\(\sigma_{\mathrm{PM},\mathrm{T}}\)) proper motion
                 velocity dispersions and proper motion anisotropy ratio
                 (\(\sigma_{\mathrm{PM},\mathrm{T}} /
                 \sigma_{\mathrm{PM},\mathrm{R}}\)), for the fit of \NGC104.
                 The dark and light shaded regions represent the \(1\sigma\)
                 and \(2\sigma\) credible intervals of the model fits,
                 respectively.
                 The observational datasets used to constrain the models are
                 shown alongside their \(1\sigma\) uncertainties by the orange
                 and green points and errorbars.
                 The models are fit only on the radial and tangential components
                 of the proper motion individually, while the total and anisotropy
                 ratio are included here solely to demonstrate the fit.
                 The background value subtracted from the number density
                 profile is shown by the dashed line.}
        \label{fig:NGC0104_kin_profiles}
    \end{figure*}
\begin{landscape}
    \begin{figure}
        \centering
        \includegraphics[width=\linewidth]{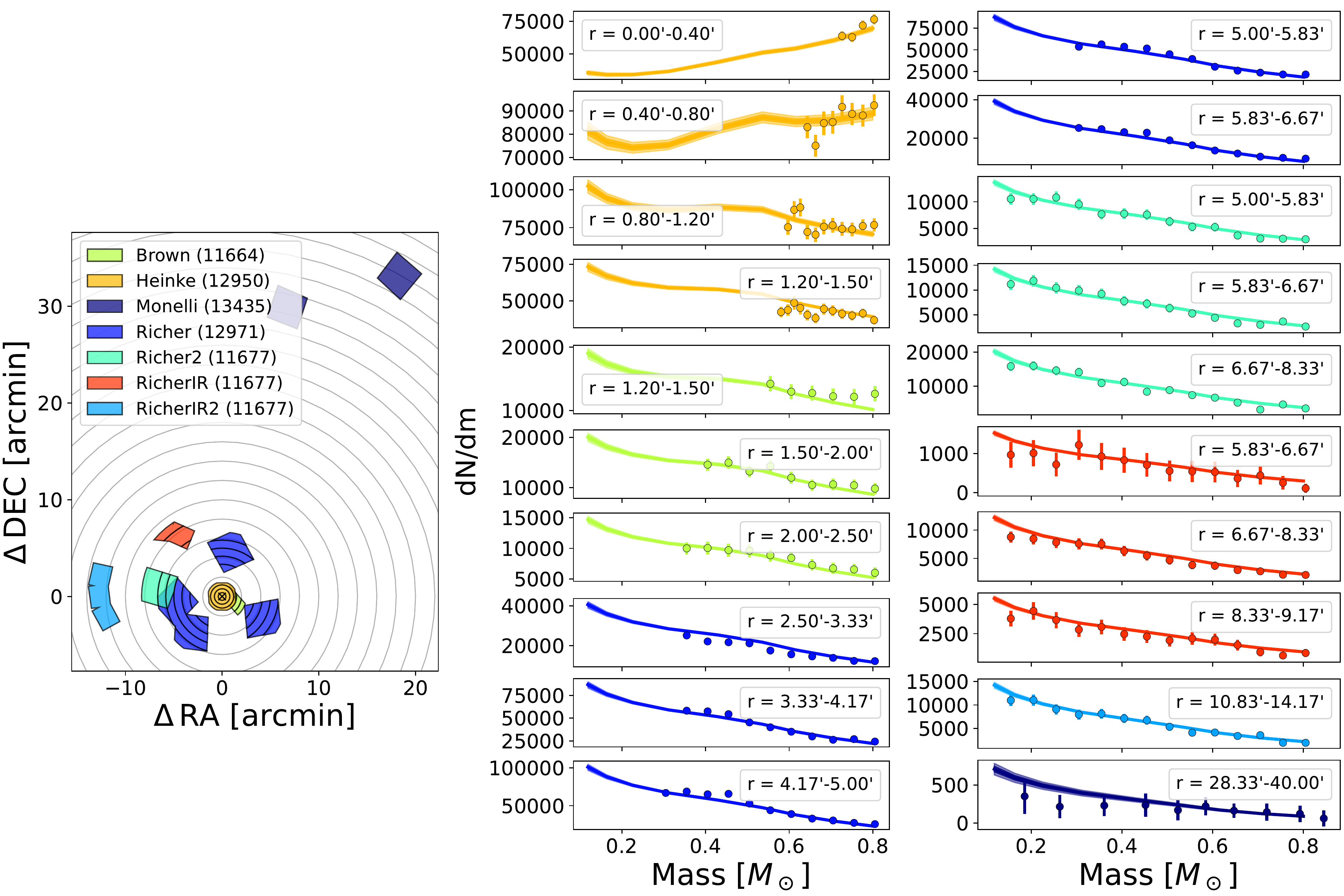}
        \caption{Model local stellar mass functions fit to the observations
                 of \NGC104. Each panel (centre and right columns) shows the
                 number of stars per unit mass
                 as a function of stellar mass, for different distance ranges
                 from the cluster centre. The dark and light shaded regions
                 represent the \(1\sigma\) and \(2\sigma\) credible intervals
                 of the model fits, respectively. The measurements used to
                 constrain the models are shown alongside their \(1\sigma\)
                 uncertainties by the points and errorbars. Each individual HST
                 observing program is denoted by a separate colour, and
                 the corresponding fields for each are shown on the left panel.
                 }
        \label{fig:NGC0104_massfunc_profiles}
    \end{figure}
    \end{landscape}

    \Cref{fig:NGC0104_kin_profiles,fig:NGC0104_massfunc_profiles} show an
    example (also for \NGC104) of the observables predicted by the best-fitting
    models, overlaid with the observational datasets used to constrain them.

    The best-fitting models for the majority of clusters match the given data
    extraordinarily well. There are, however, a small number of clusters, from
    our original sample of 37 clusters, where
    the fits do not reproduce certain datasets adequately.
    This tends to occur in systems with small amounts of PM and LOS
    velocity data. Having few kinematic datapoints, as compared
    to the mass function and number density datasets, means that these models are
    less able to constrain the non-visible mass and are prone to overfitting
    the mass functions, at the expense of the kinematics.
    As fitting both the visible and dark components
    well is vital to our analysis of the high-mass mass function and the remnant
    populations, we choose to remove these clusters from our sample going
    forward.
    Three such clusters (\NGC4590, \NGC6656, \NGC6981) were discarded due
    to their unsatisfactory fits.
    The remaining 34 clusters have best-fitting
    models that are well matched to all datasets and will make up the set of
    clusters used in all further analysis.


    \begin{landscape}
    \begin{table}
    \renewcommand*{\arraystretch}{1.4}
    \centering
    \scriptsize
    \begin{tabular}{llllllllllllll}
        \hline
        Cluster &      \multicolumn{1}{c}{$\hat{\phi}_0$} & \multicolumn{1}{c}{$M\ [10^6\ M_\odot]$} & \multicolumn{1}{c}{$r_{h}\ \left[\mathrm{pc}\right]$} & \multicolumn{1}{c}{$\log_{10}\left(\hat{r}_{a}\right)$} & \multicolumn{1}{c}{$g$} & \multicolumn{1}{c}{$\delta$} & \multicolumn{1}{c}{$\alpha_{1}$} & \multicolumn{1}{c}{$\alpha_{2}$} & \multicolumn{1}{c}{$\alpha_{3}$} & \multicolumn{1}{c}{$\mathrm{BH}_{\mathrm{ret}}\ \left[\%\right]$} & \multicolumn{1}{c}{$d\ \left[\mathrm{kpc}\right]$} &\multicolumn{1}{c}{$s^{2}\ [\mathrm{arcmin^{-4}}]$} & \multicolumn{1}{c}{$F$} \\
        \hline
        \NGC104 &       \(6.35\substack{+0.04 \\ -0.04}\) &       \(0.894\substack{+0.003 \\ -0.003}\) &    \(6.69\substack{+0.04 \\ -0.04}\) &       \(1.84\substack{+0.06 \\ -0.05}\) &    \(1.53\substack{+0.02 \\ -0.02}\) &    \(0.423\substack{+0.005 \\ -0.005}\) &    \(0.42\substack{+0.01 \\ -0.02}\) &     \(1.37\substack{+0.02 \\ -0.02}\) &       \(2.25\substack{+0.02 \\ -0.01}\) &       \(0.73\substack{+0.12 \\ -0.08}\) & \(4.416\substack{+0.009 \\ -0.009}\) &          \(0.011\substack{+0.006 \\ -0.005}\) & \(2.68\substack{+0.05 \\ -0.05}\) \\
        \NGC288 &       \(3.68\substack{+0.09 \\ -0.11}\) &       \(0.087\substack{+0.002 \\ -0.002}\) &    \(8.58\substack{+0.08 \\ -0.08}\) &          \(1.0\substack{+2.0 \\ -0.3}\) &    \(0.47\substack{+0.09 \\ -0.06}\) &    \(0.489\substack{+0.008 \\ -0.011}\) &    \(0.42\substack{+0.03 \\ -0.03}\) &     \(1.05\substack{+0.08 \\ -0.07}\) &       \(2.10\substack{+0.09 \\ -0.09}\) &          \(0.2\substack{+0.4 \\ -0.2}\) &    \(8.86\substack{+0.08 \\ -0.08}\) &             \(0.04\substack{+0.30 \\ -0.04}\) & \(1.44\substack{+0.07 \\ -0.07}\) \\
        \NGC362 &       \(5.45\substack{+0.08 \\ -0.04}\) &       \(0.278\substack{+0.001 \\ -0.002}\) &    \(3.44\substack{+0.03 \\ -0.02}\) &       \(1.41\substack{+0.06 \\ -0.05}\) &    \(1.45\substack{+0.06 \\ -0.04}\) &    \(0.484\substack{+0.005 \\ -0.013}\) &    \(0.67\substack{+0.02 \\ -0.01}\) &     \(0.78\substack{+0.02 \\ -0.02}\) &       \(3.01\substack{+0.03 \\ -0.03}\) &          \(5.0\substack{+0.2 \\ -0.3}\) &    \(8.85\substack{+0.02 \\ -0.01}\) &    \(0.00019\substack{+0.00010 \\ -0.00007}\) & \(2.52\substack{+0.05 \\ -0.04}\) \\
        \NGC1261 &          \(3.4\substack{+0.1 \\ -0.1}\) &       \(0.182\substack{+0.005 \\ -0.005}\) &    \(5.00\substack{+0.07 \\ -0.08}\) &          \(1.0\substack{+1.8 \\ -0.2}\) &       \(2.3\substack{+0.1 \\ -0.1}\) &       \(0.48\substack{+0.01 \\ -0.02}\) &    \(1.01\substack{+0.08 \\ -0.08}\) &     \(1.37\substack{+0.10 \\ -0.09}\) &          \(2.7\substack{+0.2 \\ -0.1}\) &                \(2\substack{+1 \\ -1}\) &      \(16.3\substack{+0.2 \\ -0.2}\) & \(0.000006\substack{+0.000014 \\ -0.000004}\) &    \(4.2\substack{+0.3 \\ -0.3}\) \\
        \NGC1851 &       \(5.71\substack{+0.04 \\ -0.03}\) &       \(0.326\substack{+0.001 \\ -0.002}\) &    \(3.68\substack{+0.02 \\ -0.02}\) &          \(4.2\substack{+0.6 \\ -1.0}\) &    \(1.84\substack{+0.01 \\ -0.02}\) &    \(0.491\substack{+0.007 \\ -0.005}\) &    \(0.59\substack{+0.02 \\ -0.02}\) &     \(1.25\substack{+0.04 \\ -0.04}\) &       \(2.88\substack{+0.05 \\ -0.05}\) &          \(2.6\substack{+0.8 \\ -0.4}\) &   \(12.22\substack{+0.05 \\ -0.07}\) & \(0.000020\substack{+0.000016 \\ -0.000007}\) &    \(2.3\substack{+0.1 \\ -0.1}\) \\
        \NGC2808 &       \(4.87\substack{+0.01 \\ -0.01}\) &       \(0.973\substack{+0.004 \\ -0.004}\) & \(3.959\substack{+0.006 \\ -0.007}\) &       \(1.50\substack{+0.04 \\ -0.05}\) & \(1.848\substack{+0.007 \\ -0.011}\) &    \(0.495\substack{+0.002 \\ -0.002}\) &    \(0.36\substack{+0.01 \\ -0.01}\) &     \(1.59\substack{+0.01 \\ -0.02}\) &       \(3.12\substack{+0.04 \\ -0.03}\) &         \(11.7\substack{+0.8 \\ -0.4}\) &   \(10.39\substack{+0.03 \\ -0.03}\) & \(0.000013\substack{+0.000004 \\ -0.000002}\) & \(2.75\substack{+0.08 \\ -0.11}\) \\
        \NGC3201 &          \(4.8\substack{+0.2 \\ -0.2}\) &       \(0.180\substack{+0.006 \\ -0.006}\) &       \(9.9\substack{+0.5 \\ -0.6}\) &       \(0.94\substack{+0.06 \\ -0.05}\) &       \(1.6\substack{+0.1 \\ -0.1}\) &       \(0.45\substack{+0.03 \\ -0.03}\) &    \(1.02\substack{+0.07 \\ -0.08}\) &     \(1.38\substack{+0.10 \\ -0.10}\) &       \(2.15\substack{+0.08 \\ -0.07}\) &       \(0.06\substack{+0.10 \\ -0.04}\) &    \(4.65\substack{+0.03 \\ -0.03}\) &               \(19.0\substack{+0.7 \\ -1.1}\) &    \(2.6\substack{+0.3 \\ -0.2}\) \\
        \NGC5024 &          \(5.1\substack{+0.1 \\ -0.1}\) &          \(0.53\substack{+0.01 \\ -0.01}\) &      \(10.5\substack{+0.2 \\ -0.2}\) &                \(7\substack{+2 \\ -2}\) &    \(2.24\substack{+0.09 \\ -0.09}\) &       \(0.32\substack{+0.03 \\ -0.02}\) &    \(0.87\substack{+0.06 \\ -0.06}\) &     \(1.59\substack{+0.06 \\ -0.06}\) &          \(2.4\substack{+0.1 \\ -0.1}\) &                \(2\substack{+3 \\ -1}\) &      \(18.3\substack{+0.1 \\ -0.2}\) &          \(0.002\substack{+0.003 \\ -0.001}\) &    \(2.7\substack{+0.2 \\ -0.1}\) \\
        \NGC5139 &       \(2.44\substack{+0.13 \\ -0.07}\) &          \(3.21\substack{+0.04 \\ -0.03}\) &    \(9.62\substack{+0.06 \\ -0.04}\) &          \(6.4\substack{+0.4 \\ -0.7}\) &    \(2.53\substack{+0.02 \\ -0.02}\) &    \(0.426\substack{+0.006 \\ -0.005}\) & \(0.830\substack{+0.009 \\ -0.008}\) &     \(1.19\substack{+0.05 \\ -0.02}\) &       \(2.16\substack{+0.04 \\ -0.08}\) &               \(20\substack{+3 \\ -5}\) &    \(5.35\substack{+0.02 \\ -0.02}\) &    \(0.00027\substack{+0.00006 \\ -0.00006}\) &    \(6.2\substack{+0.2 \\ -0.1}\) \\
        \NGC5272 &       \(5.67\substack{+0.05 \\ -0.02}\) &       \(0.488\substack{+0.004 \\ -0.007}\) &    \(7.04\substack{+0.06 \\ -0.05}\) &                \(5\substack{+2 \\ -1}\) &    \(1.84\substack{+0.03 \\ -0.03}\) &    \(0.306\substack{+0.002 \\ -0.003}\) &    \(1.06\substack{+0.02 \\ -0.02}\) &     \(1.29\substack{+0.04 \\ -0.03}\) &       \(2.24\substack{+0.06 \\ -0.05}\) &          \(1.8\substack{+0.6 \\ -0.4}\) &   \(10.13\substack{+0.06 \\ -0.06}\) & \(0.000010\substack{+0.000043 \\ -0.000008}\) & \(2.10\substack{+0.07 \\ -0.08}\) \\
        \NGC5904 &          \(5.4\substack{+0.1 \\ -0.1}\) &       \(0.385\substack{+0.006 \\ -0.007}\) &    \(6.35\substack{+0.08 \\ -0.08}\) &                \(5\substack{+2 \\ -2}\) &    \(1.44\substack{+0.06 \\ -0.06}\) &       \(0.45\substack{+0.03 \\ -0.03}\) &    \(0.48\substack{+0.04 \\ -0.04}\) &     \(0.79\substack{+0.08 \\ -0.08}\) &       \(2.27\substack{+0.06 \\ -0.06}\) &          \(0.2\substack{+0.3 \\ -0.2}\) &    \(7.36\substack{+0.05 \\ -0.05}\) &          \(0.006\substack{+0.010 \\ -0.004}\) &    \(4.2\substack{+0.2 \\ -0.2}\) \\
        \NGC5986 &       \(4.12\substack{+0.05 \\ -0.06}\) &       \(0.294\substack{+0.008 \\ -0.006}\) &    \(4.41\substack{+0.04 \\ -0.03}\) &          \(1.0\substack{+2.2 \\ -0.1}\) &    \(1.38\substack{+0.07 \\ -0.05}\) &    \(0.492\substack{+0.005 \\ -0.013}\) &    \(0.50\substack{+0.03 \\ -0.03}\) &     \(0.99\substack{+0.04 \\ -0.04}\) &       \(2.41\substack{+0.07 \\ -0.08}\) &       \(0.13\substack{+0.11 \\ -0.08}\) &   \(10.31\substack{+0.10 \\ -0.09}\) &    \(0.00013\substack{+0.00008 \\ -0.00008}\) &    \(2.2\substack{+0.2 \\ -0.1}\) \\
        \NGC6093 &       \(6.23\substack{+0.07 \\ -0.08}\) &       \(0.302\substack{+0.004 \\ -0.004}\) &    \(2.38\substack{+0.02 \\ -0.02}\) &                \(6\substack{+2 \\ -2}\) &    \(1.40\substack{+0.05 \\ -0.06}\) &       \(0.34\substack{+0.02 \\ -0.02}\) &    \(0.03\substack{+0.07 \\ -0.07}\) &     \(0.96\substack{+0.10 \\ -0.11}\) &       \(2.48\substack{+0.05 \\ -0.05}\) &          \(4.5\substack{+0.4 \\ -0.6}\) &    \(9.99\substack{+0.07 \\ -0.07}\) &          \(0.009\substack{+0.003 \\ -0.002}\) &    \(3.7\substack{+0.3 \\ -0.3}\) \\
        \NGC6121 &          \(6.4\substack{+0.2 \\ -0.1}\) &       \(0.090\substack{+0.002 \\ -0.001}\) &    \(3.90\substack{+0.06 \\ -0.05}\) &                \(7\substack{+2 \\ -3}\) &    \(0.87\substack{+0.07 \\ -0.08}\) &       \(0.46\substack{+0.02 \\ -0.03}\) &   \(-0.10\substack{+0.08 \\ -0.08}\) &     \(0.34\substack{+0.08 \\ -0.08}\) &       \(2.26\substack{+0.06 \\ -0.06}\) &          \(0.4\substack{+0.2 \\ -0.2}\) &    \(1.85\substack{+0.01 \\ -0.01}\) &    \(0.00001\substack{+0.00006 \\ -0.00001}\) & \(1.28\substack{+0.08 \\ -0.07}\) \\
        \NGC6171 &       \(5.57\substack{+0.08 \\ -0.06}\) &       \(0.063\substack{+0.002 \\ -0.002}\) &    \(3.89\substack{+0.05 \\ -0.05}\) &                \(5\substack{+2 \\ -2}\) &    \(0.35\substack{+0.09 \\ -0.09}\) &    \(0.487\substack{+0.010 \\ -0.017}\) &   \(-0.07\substack{+0.03 \\ -0.03}\) & \(-0.004\substack{+0.042 \\ -0.037}\) &       \(2.34\substack{+0.07 \\ -0.07}\) &          \(0.4\substack{+0.3 \\ -0.2}\) &    \(5.60\substack{+0.06 \\ -0.07}\) &          \(0.004\substack{+0.002 \\ -0.002}\) & \(1.46\substack{+0.10 \\ -0.09}\) \\
        \NGC6205 &          \(3.5\substack{+0.1 \\ -0.1}\) &       \(0.425\substack{+0.008 \\ -0.008}\) &    \(4.51\substack{+0.04 \\ -0.04}\) &                \(6\substack{+3 \\ -3}\) &    \(2.53\substack{+0.06 \\ -0.06}\) &       \(0.41\substack{+0.04 \\ -0.05}\) &       \(0.3\substack{+0.1 \\ -0.1}\) &     \(0.96\substack{+0.08 \\ -0.08}\) &       \(2.38\substack{+0.08 \\ -0.08}\) &                \(4\substack{+2 \\ -1}\) &    \(7.33\substack{+0.06 \\ -0.05}\) &    \(0.00002\substack{+0.00003 \\ -0.00001}\) &    \(2.9\substack{+0.4 \\ -0.3}\) \\
        \NGC6218 &       \(5.09\substack{+0.04 \\ -0.05}\) &       \(0.100\substack{+0.002 \\ -0.001}\) &    \(4.13\substack{+0.04 \\ -0.04}\) &                \(4\substack{+2 \\ -2}\) &    \(0.53\substack{+0.06 \\ -0.06}\) &    \(0.496\substack{+0.003 \\ -0.005}\) &    \(0.11\substack{+0.03 \\ -0.04}\) &     \(0.18\substack{+0.05 \\ -0.04}\) &       \(2.81\substack{+0.05 \\ -0.06}\) &          \(3.6\substack{+0.6 \\ -0.7}\) &    \(5.03\substack{+0.04 \\ -0.04}\) &    \(0.00008\substack{+0.00011 \\ -0.00006}\) &    \(2.2\substack{+0.2 \\ -0.2}\) \\
        \NGC6254 &          \(5.7\substack{+0.2 \\ -0.2}\) &       \(0.211\substack{+0.004 \\ -0.004}\) &    \(5.18\substack{+0.07 \\ -0.07}\) &                \(3\substack{+1 \\ -1}\) &    \(0.97\substack{+0.08 \\ -0.09}\) &       \(0.41\substack{+0.04 \\ -0.04}\) &    \(0.32\substack{+0.04 \\ -0.05}\) &     \(0.91\substack{+0.08 \\ -0.08}\) &       \(2.20\substack{+0.07 \\ -0.06}\) &          \(0.3\substack{+0.4 \\ -0.2}\) &    \(5.10\substack{+0.04 \\ -0.04}\) &       \(0.0003\substack{+0.0002 \\ -0.0002}\) &    \(2.7\substack{+0.2 \\ -0.2}\) \\
        \NGC6266 &          \(5.6\substack{+0.1 \\ -0.2}\) &          \(0.75\substack{+0.02 \\ -0.01}\) &    \(3.17\substack{+0.07 \\ -0.06}\) &       \(1.38\substack{+0.12 \\ -0.07}\) &    \(0.91\substack{+0.06 \\ -0.05}\) &    \(0.498\substack{+0.002 \\ -0.003}\) &       \(0.2\substack{+0.2 \\ -0.2}\) &     \(1.24\substack{+0.07 \\ -0.06}\) &       \(2.24\substack{+0.03 \\ -0.04}\) &                 \multicolumn{1}{c}{---} &    \(6.50\substack{+0.04 \\ -0.03}\) &             \(0.22\substack{+0.02 \\ -0.02}\) &    \(2.5\substack{+0.3 \\ -0.3}\) \\
        \NGC6341 &          \(5.4\substack{+0.4 \\ -0.3}\) &       \(0.300\substack{+0.007 \\ -0.005}\) &    \(4.20\substack{+0.06 \\ -0.07}\) &                \(6\substack{+3 \\ -3}\) &    \(1.78\substack{+0.05 \\ -0.06}\) &       \(0.43\substack{+0.04 \\ -0.05}\) &    \(0.81\substack{+0.04 \\ -0.04}\) &     \(1.09\substack{+0.09 \\ -0.09}\) &       \(2.21\substack{+0.08 \\ -0.08}\) &          \(0.7\substack{+0.3 \\ -0.4}\) &    \(8.42\substack{+0.06 \\ -0.06}\) &    \(0.00007\substack{+0.00007 \\ -0.00004}\) &    \(3.7\substack{+0.2 \\ -0.2}\) \\
        \NGC6352 &          \(6.6\substack{+0.2 \\ -0.1}\) &       \(0.098\substack{+0.006 \\ -0.006}\) &       \(8.7\substack{+0.7 \\ -0.7}\) &                \(7\substack{+2 \\ -2}\) &       \(0.2\substack{+0.2 \\ -0.1}\) &    \(0.491\substack{+0.007 \\ -0.013}\) &    \(0.25\substack{+0.10 \\ -0.10}\) &     \(0.66\substack{+0.09 \\ -0.10}\) &       \(1.99\substack{+0.05 \\ -0.06}\) &       \(0.16\substack{+0.04 \\ -0.04}\) &    \(5.62\substack{+0.06 \\ -0.06}\) &             \(2.04\substack{+0.05 \\ -0.03}\) &    \(2.1\substack{+0.2 \\ -0.1}\) \\
        \NGC6362 &       \(4.14\substack{+0.13 \\ -0.10}\) &       \(0.111\substack{+0.003 \\ -0.003}\) &    \(6.95\substack{+0.09 \\ -0.09}\) &                \(4\substack{+2 \\ -2}\) &       \(0.5\substack{+0.2 \\ -0.2}\) &       \(0.46\substack{+0.02 \\ -0.04}\) &    \(0.26\substack{+0.04 \\ -0.04}\) &     \(0.64\substack{+0.07 \\ -0.07}\) &       \(1.84\substack{+0.07 \\ -0.07}\) &       \(0.03\substack{+0.05 \\ -0.03}\) &    \(7.63\substack{+0.06 \\ -0.06}\) &                     \(16\substack{+3 \\ -4}\) & \(1.57\substack{+0.11 \\ -0.09}\) \\
        \NGC6366 &          \(4.0\substack{+0.1 \\ -0.1}\) &       \(0.032\substack{+0.001 \\ -0.001}\) &    \(4.62\substack{+0.09 \\ -0.09}\) &                \(2\substack{+2 \\ -1}\) &       \(0.6\substack{+0.2 \\ -0.2}\) &    \(0.488\substack{+0.009 \\ -0.017}\) &   \(-0.30\substack{+0.07 \\ -0.08}\) &    \(-0.19\substack{+0.08 \\ -0.07}\) &          \(3.1\substack{+0.2 \\ -0.2}\) &                \(4\substack{+5 \\ -3}\) &    \(3.38\substack{+0.04 \\ -0.04}\) &          \(0.002\substack{+0.002 \\ -0.001}\) & \(1.25\substack{+0.09 \\ -0.08}\) \\
        \NGC6397 &       \(7.86\substack{+0.09 \\ -0.09}\) &       \(0.108\substack{+0.002 \\ -0.002}\) &       \(4.9\substack{+0.1 \\ -0.1}\) &          \(2.7\substack{+0.1 \\ -0.1}\) &    \(1.58\substack{+0.07 \\ -0.07}\) &    \(0.497\substack{+0.002 \\ -0.003}\) &    \(0.60\substack{+0.04 \\ -0.04}\) &     \(0.69\substack{+0.07 \\ -0.05}\) &       \(2.42\substack{+0.04 \\ -0.03}\) &       \(0.14\substack{+0.01 \\ -0.01}\) &    \(2.43\substack{+0.01 \\ -0.01}\) &                \(8.8\substack{+0.8 \\ -1.2}\) &    \(2.1\substack{+0.1 \\ -0.1}\) \\
        \NGC6541 &       \(6.03\substack{+0.06 \\ -0.04}\) &       \(0.220\substack{+0.002 \\ -0.002}\) &    \(3.49\substack{+0.03 \\ -0.02}\) &       \(3.85\substack{+0.09 \\ -0.07}\) &    \(1.33\substack{+0.01 \\ -0.02}\) &    \(0.470\substack{+0.002 \\ -0.003}\) &    \(0.30\substack{+0.01 \\ -0.01}\) &     \(1.18\substack{+0.05 \\ -0.05}\) &    \(2.107\substack{+0.009 \\ -0.009}\) &    \(0.006\substack{+0.007 \\ -0.005}\) & \(7.491\substack{+0.009 \\ -0.011}\) &       \(0.0003\substack{+0.0002 \\ -0.0001}\) & \(4.45\substack{+0.08 \\ -0.08}\) \\
        \NGC6624 &         \(11.0\substack{+0.2 \\ -0.2}\) &       \(0.102\substack{+0.003 \\ -0.003}\) &    \(2.13\substack{+0.05 \\ -0.04}\) &       \(2.91\substack{+0.07 \\ -0.07}\) &       \(1.2\substack{+0.1 \\ -0.1}\) &    \(0.496\substack{+0.003 \\ -0.004}\) &   \(-0.70\substack{+0.09 \\ -0.09}\) &    \(-0.50\substack{+0.05 \\ -0.05}\) &       \(2.40\substack{+0.05 \\ -0.06}\) &                 \multicolumn{1}{c}{---} &    \(8.11\substack{+0.08 \\ -0.06}\) &                \(0.3\substack{+0.1 \\ -0.1}\) &    \(1.2\substack{+0.1 \\ -0.1}\) \\
        \NGC6681 &       \(7.28\substack{+0.08 \\ -0.06}\) &       \(0.096\substack{+0.002 \\ -0.002}\) &    \(2.67\substack{+0.05 \\ -0.05}\) &       \(1.82\substack{+0.10 \\ -0.07}\) &    \(1.16\substack{+0.06 \\ -0.08}\) &    \(0.487\substack{+0.009 \\ -0.009}\) &   \(-0.13\substack{+0.08 \\ -0.09}\) &     \(0.13\substack{+0.08 \\ -0.10}\) &       \(2.04\substack{+0.05 \\ -0.04}\) &       \(0.10\substack{+0.02 \\ -0.01}\) &    \(9.46\substack{+0.04 \\ -0.05}\) &             \(0.05\substack{+0.01 \\ -0.01}\) &    \(3.2\substack{+0.2 \\ -0.2}\) \\
        \NGC6723 &       \(3.74\substack{+0.92 \\ -0.08}\) &       \(0.177\substack{+0.017 \\ -0.002}\) &    \(5.12\substack{+0.05 \\ -0.03}\) &       \(0.35\substack{+2.57 \\ -0.02}\) &    \(0.59\substack{+0.65 \\ -0.07}\) &       \(0.31\substack{+0.09 \\ -0.01}\) &    \(0.10\substack{+0.02 \\ -0.02}\) &     \(0.34\substack{+0.06 \\ -0.03}\) &    \(2.360\substack{+0.009 \\ -0.328}\) &          \(4.6\substack{+0.5 \\ -4.5}\) & \(8.388\substack{+0.068 \\ -0.008}\) &               \(13.6\substack{+0.6 \\ -0.3}\) & \(2.36\substack{+0.06 \\ -0.80}\) \\
        \NGC6752 &       \(6.07\substack{+0.06 \\ -0.05}\) &       \(0.211\substack{+0.004 \\ -0.003}\) &    \(3.54\substack{+0.04 \\ -0.04}\) &       \(1.54\substack{+0.02 \\ -0.02}\) &    \(1.64\substack{+0.02 \\ -0.02}\) &    \(0.496\substack{+0.003 \\ -0.004}\) &    \(0.45\substack{+0.05 \\ -0.06}\) &     \(0.51\substack{+0.04 \\ -0.04}\) &       \(2.08\substack{+0.04 \\ -0.04}\) &                 \multicolumn{1}{c}{---} &    \(4.04\substack{+0.03 \\ -0.02}\) &    \(0.00003\substack{+0.00003 \\ -0.00002}\) &    \(5.1\substack{+0.2 \\ -0.2}\) \\
        \NGC6779 &       \(5.88\substack{+0.05 \\ -0.05}\) &       \(0.168\substack{+0.004 \\ -0.004}\) &    \(5.38\substack{+0.07 \\ -0.07}\) &          \(2.8\substack{+0.8 \\ -0.6}\) &    \(1.00\substack{+0.10 \\ -0.10}\) &    \(0.314\substack{+0.010 \\ -0.008}\) &    \(0.57\substack{+0.05 \\ -0.05}\) &     \(0.86\substack{+0.07 \\ -0.07}\) &       \(2.30\substack{+0.06 \\ -0.06}\) &          \(3.4\substack{+0.6 \\ -0.6}\) &   \(10.13\substack{+0.04 \\ -0.05}\) &       \(0.0003\substack{+0.0002 \\ -0.0002}\) &    \(2.9\substack{+0.1 \\ -0.1}\) \\
        \NGC6809 &       \(3.09\substack{+0.07 \\ -0.06}\) &       \(0.183\substack{+0.005 \\ -0.004}\) &    \(6.18\substack{+0.05 \\ -0.05}\) &                \(6\substack{+3 \\ -3}\) &    \(1.26\substack{+0.07 \\ -0.08}\) &    \(0.488\substack{+0.009 \\ -0.015}\) &    \(0.50\substack{+0.03 \\ -0.03}\) &     \(0.58\substack{+0.05 \\ -0.05}\) &          \(2.6\substack{+0.1 \\ -0.1}\) &          \(0.7\substack{+0.5 \\ -0.4}\) &    \(5.25\substack{+0.03 \\ -0.04}\) &       \(0.0005\substack{+0.0005 \\ -0.0003}\) &    \(2.8\substack{+0.2 \\ -0.2}\) \\
        \NGC7078 &          \(8.8\substack{+0.2 \\ -0.2}\) &       \(0.614\substack{+0.007 \\ -0.007}\) &    \(5.17\substack{+0.05 \\ -0.04}\) &       \(3.02\substack{+0.08 \\ -0.09}\) &    \(1.61\substack{+0.07 \\ -0.09}\) &       \(0.46\substack{+0.01 \\ -0.01}\) &    \(0.80\substack{+0.04 \\ -0.04}\) &     \(1.75\substack{+0.06 \\ -0.06}\) &       \(1.94\substack{+0.04 \\ -0.04}\) &                 \multicolumn{1}{c}{---} &   \(10.69\substack{+0.06 \\ -0.06}\) &    \(0.00004\substack{+0.00003 \\ -0.00002}\) &    \(3.1\substack{+0.2 \\ -0.2}\) \\
        \NGC7089 &          \(4.9\substack{+0.2 \\ -0.2}\) &       \(0.627\substack{+0.009 \\ -0.009}\) &    \(4.59\substack{+0.07 \\ -0.07}\) &                \(9\substack{+4 \\ -4}\) &    \(1.92\substack{+0.06 \\ -0.07}\) &       \(0.43\substack{+0.04 \\ -0.04}\) &       \(0.5\substack{+0.1 \\ -0.1}\) &     \(1.14\substack{+0.07 \\ -0.07}\) &       \(2.97\substack{+0.08 \\ -0.08}\) &               \(12\substack{+4 \\ -4}\) &   \(11.53\substack{+0.08 \\ -0.07}\) &    \(0.00006\substack{+0.00005 \\ -0.00003}\) &    \(3.0\substack{+0.3 \\ -0.3}\) \\
        \NGC7099 & \(6.3645\substack{+0.0004 \\ -0.0003}\) & \(0.14210\substack{+0.00003 \\ -0.00004}\) & \(4.529\substack{+0.005 \\ -0.011}\) & \(1.6758\substack{+0.0009 \\ -0.0019}\) & \(1.172\substack{+0.003 \\ -0.008}\) & \(0.4931\substack{+0.0001 \\ -0.0003}\) & \(0.499\substack{+0.002 \\ -0.001}\) &  \(1.169\substack{+0.003 \\ -0.006}\) & \(2.2312\substack{+0.0004 \\ -0.0009}\) & \(0.1152\substack{+0.0001 \\ -0.0001}\) & \(8.445\substack{+0.002 \\ -0.001}\) &    \(0.00005\substack{+0.00002 \\ -0.00003}\) & \(3.65\substack{+0.03 \\ -0.01}\) \\
        \hline
    \end{tabular}
    \caption{Median and \(1\sigma\) uncertainties of all best-fitting
             model, mass function and nuisance parameters, for all
             clusters. Empty dashes
             in the \(\BHret\) column represent the four
             core-collapsed clusters with a value fixed to 0.
             Note that all uncertainties presented here represent only
             the statistical uncertainties on the fits, and likely
             underestimate the true uncertainties.
             See \cref{sec:cluster_parameters,sub:high_mass_imf} for more
             details.}
    \label{table:best_fitting_params}
    \end{table}
    \end{landscape}

\subsection{Cluster Parameters}\label{sec:cluster_parameters}

    Given this set of best-fitting models, we next examine the distributions
    of various model parameters and compare with other results from the
    literature. The best-fitting model, mass function and nuisance
    parameters for all clusters are shown in \Cref{table:best_fitting_params}.

    It must be noted here that all uncertainties presented for these parameters,
    in this entire analysis, are accounting solely for the statistical
    uncertainties on the parameter fits. Our fitting procedure operates under
    the assumption that our models are a good representation of
    the data, and as such may, in reality, be underestimating the true errors.
    It has been shown that multimass DF models, such as those used here, may
    underestimate the uncertainties when compared to more flexible models,
    such as Jeans models \citep{VHB2019}, which could
    be indicative of systematic errors not captured in the statistical
    uncertainties and limitations in the ability of these models to perfectly
    reproduce the data.

\subsubsection{Comparison with literature}

\begin{figure*}
        \centering
        \includegraphics[width=\linewidth]{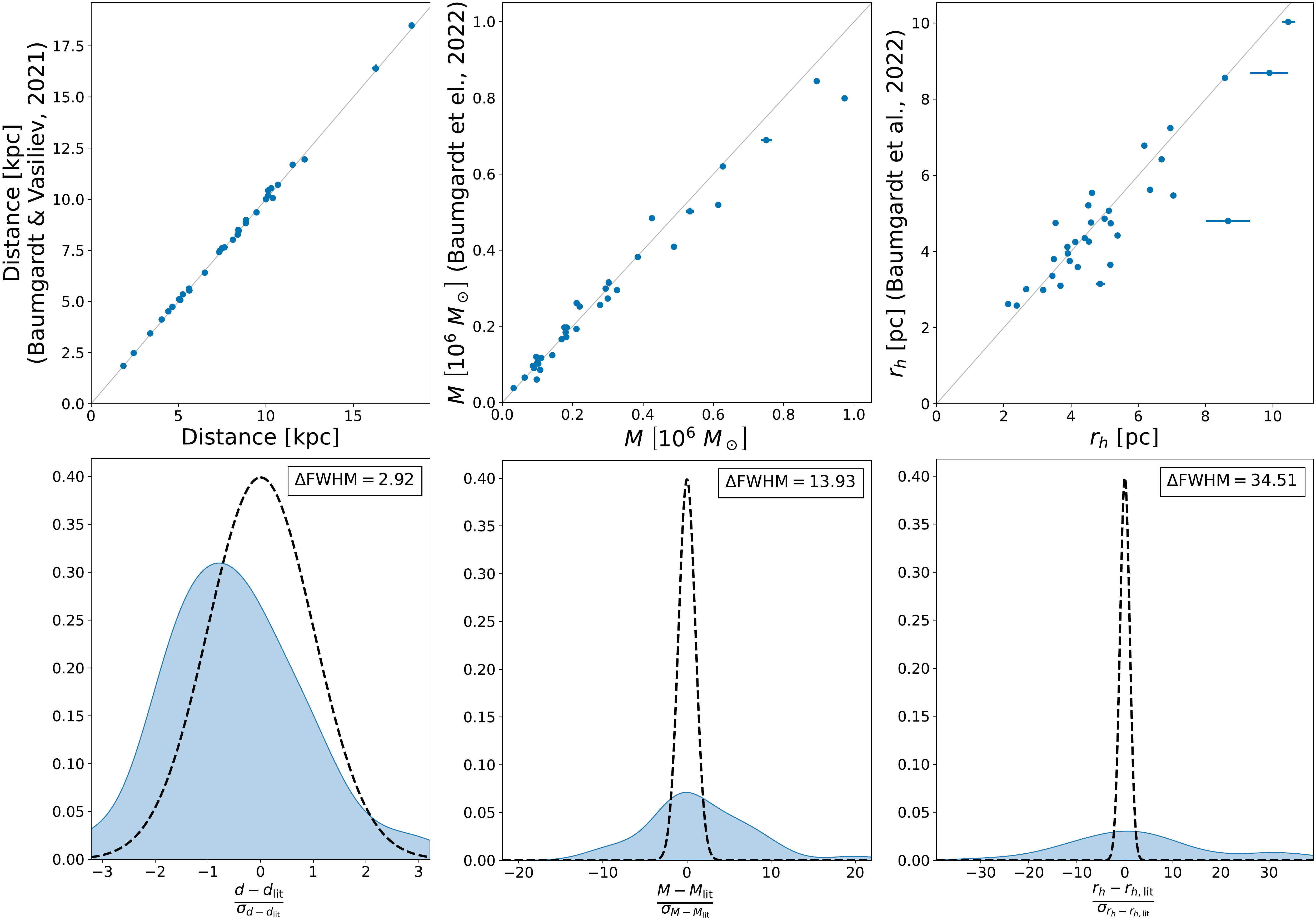}
        \caption{Comparison of the heliocentric distance, total system mass, and
                 half-mass radius of all cluster fits against the distances
                 computed by \citet{Vasiliev2021} and the properties inferred
                 from the \Nbody model fits of \citet{Baumgardt2022}.
                 The top row compares the median and \(1\sigma\) values of
                 both both sources, with the grey diagonal representing
                 perfect agreement. The bottom row shows the distribution
                 (represented by a Gaussian KDE) of the fractional
                 differences among all clusters, divided by their
                 combined uncertainties. The dashed black line shows
                 a Gaussian, centred on 0 with a width of \(\sigma=1\), which
                 represents perfect agreement. The ratio of the FWHM of the
                 fractional difference distributions to that of the Gaussian
                 is noted in the top right corners.
                 \NGC5139 is excluded from these figures due to its very
                 large mass but shows similar agreement, and will be
                 discussed in more detail in \paperII.}
        \label{fig:model_comps}
    \end{figure*}

    To begin, we can compare our best-fitting models with other
    comprehensive studies of Milky Way GCs in the literature.
    Namely, we consider in \Cref{fig:model_comps} the distances determined by
    \citet{Vasiliev2021}, and the total masses and half-mass radii inferred
    from the \Nbody model fits of \citet{Baumgardt2022}.
    The top panels of \Cref{fig:model_comps} shows the relation between the
    literature values and our own, demonstrating a good general agreement
    and no obvious biases,
    but with noticeable spread around the lines of perfect agreement.
    This is reinforced in the bottom panels, which
    show the distribution of the fractional differences between our values,
    divided by their combined uncertainties. If the agreement was perfect,
    and all systematic errors were accounted for in our uncertainties,
    this distribution would resemble a Gaussian centred at 0 with a
    width of \(\sigma=1\) (shown by the dashed black line). Our
    distributions, meanwhile, are roughly centred near 0 but, in the
    case of mass and half-mass radius, show a much wider spread.
    This increased width demonstrates that the combined systematic
    uncertainties of both our values and the literature values are
    underestimated by a factor of a few.

    It should be noted that the excellent agreement with the heliocentric
    distances of \citet{Vasiliev2021} is largely by design, given our use of
    a Gaussian centred on their values as the prior on the distance parameter
    (see \Cref{subsub:priors}), but this does still indicate that our models
    and observations are perfectly compatible with those distances.

\subsubsection{Anisotropy}\label{sub:anisotropy}

    \begin{figure*}
        \centering
        \includegraphics[width=\linewidth]{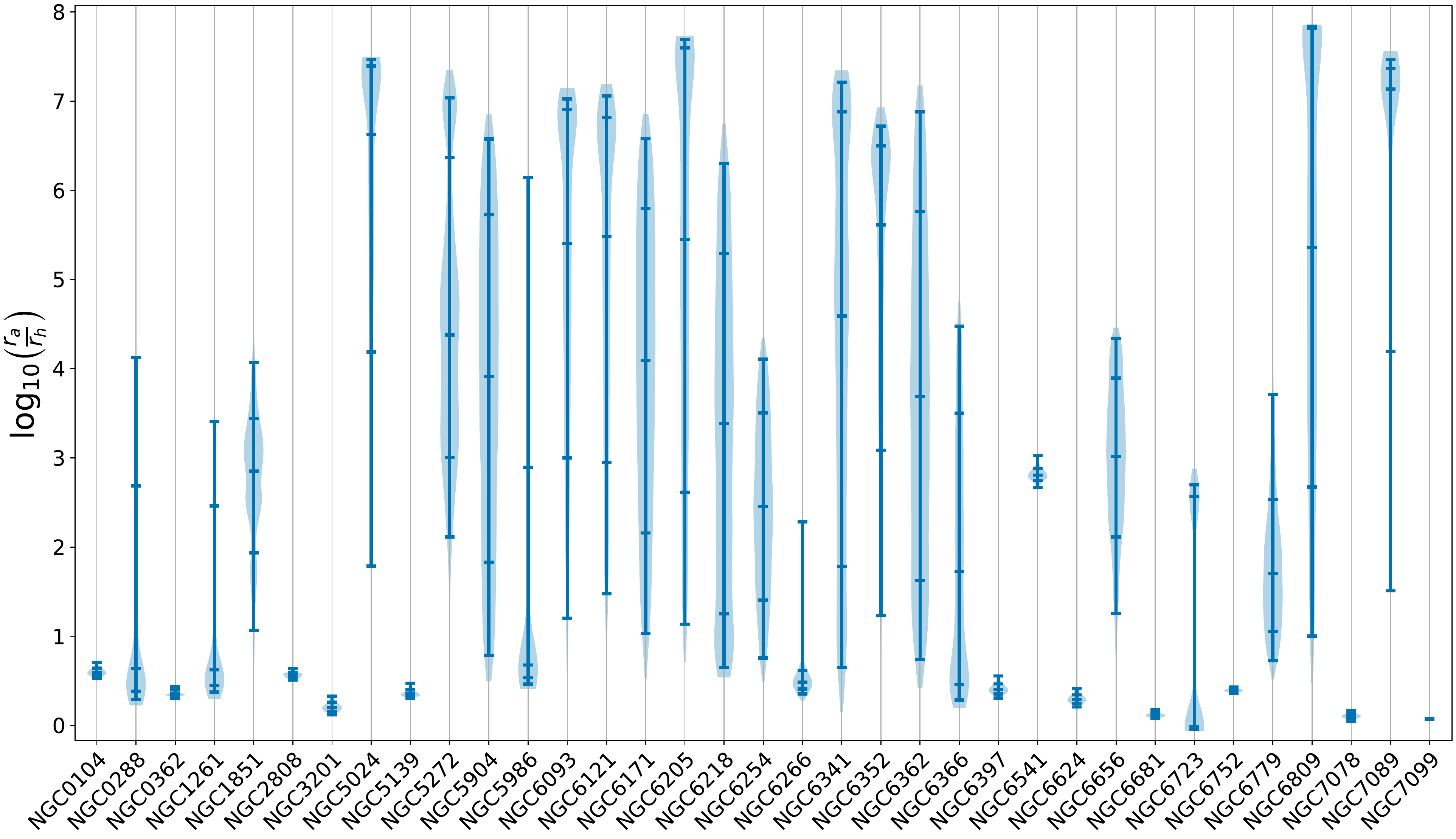}
        \caption{Violin plot of the posterior distribution of the
                 (log) anisotropy radius, normalized by the cluster
                 half-mass radius, for all clusters.
                 The median, \(1\sigma\) and \(2\sigma\) values are
                 denoted by the horizontal ticks on each distribution.}
        \label{fig:ra_distributions}
    \end{figure*}

    \Cref{fig:ra_distributions} shows the distribution of the anisotropy radius
    (normalized by the half-mass radius) for all clusters. It is immediately
    clear from this plot that
    there are two populations of anisotropy results in our fits; distributions
    with a clear peak, constrained to a narrow range of best-fitting values,
    and very broad, flat distributions with no clear peak above a certain
    minimum value, extending up to the prior bounds.
    As described in \Cref{sub:model_parameters}, clusters where \(\ra\)
    is more clearly peaked favour a certain amount of radial anisotropy,
    whereas clusters with broad posterior distributions are effectively
    isotropic, as all values of \(\ra\) above the minima of the
    broad distributions (corresponding to approximately at or above the
    truncation radius) essentially lead to the same isotropic model.
    Values above this minimum
    therefore have a negligible effect on the computed model likelihoods.

    It should be noted that the constraints we can place on velocity anisotropy
    come entirely from the Gaia and HST
    proper motion dispersion profiles. These datasets are quite limited in many
    clusters, and as such some of the clusters with broad distributions may
    not actually be entirely isotropic in reality, but simply cannot be
    sufficiently constrained by the data currently available.
    It is also important to note that our \limepy models are unable to
    reproduce any amount of tangential anisotropy \citep{Gieles2015},
    and instead, when tangentially biased anisotropy is present in our data.
    the models will favour a mostly isotropic fit as a compromise between the
    radial and tangential regimes \citep{Peuten2017}.

    It is clear, based on the wide range of \(\log(\ra)\) values, that allowing
    the anisotropy radius to vary freely is necessary to best model the GCs.
    The degree of anisotropy in a cluster is important for understanding the
    central dark remnant populations, as there exists a degeneracy between
    the observational fingerprints of central dark mass and velocity anisotropy,
    when observational constraints on only one component of velocity are
    available \citep{Evans2009, Zocchi2017}.


\section{Mass Functions}\label{ch:mass_functions}


    In this section, we explore the mass function exponents inferred from our
    model fits. We will examine the relationships between
    the MF exponents, discuss the connection between the global mass
    function probed by our models and the initial stellar mass function of GCs,
    and search for any possible correlations between the IMF and environment of
    globular clusters.


    \begin{figure*}
        \centering
        \includegraphics[width=\linewidth]{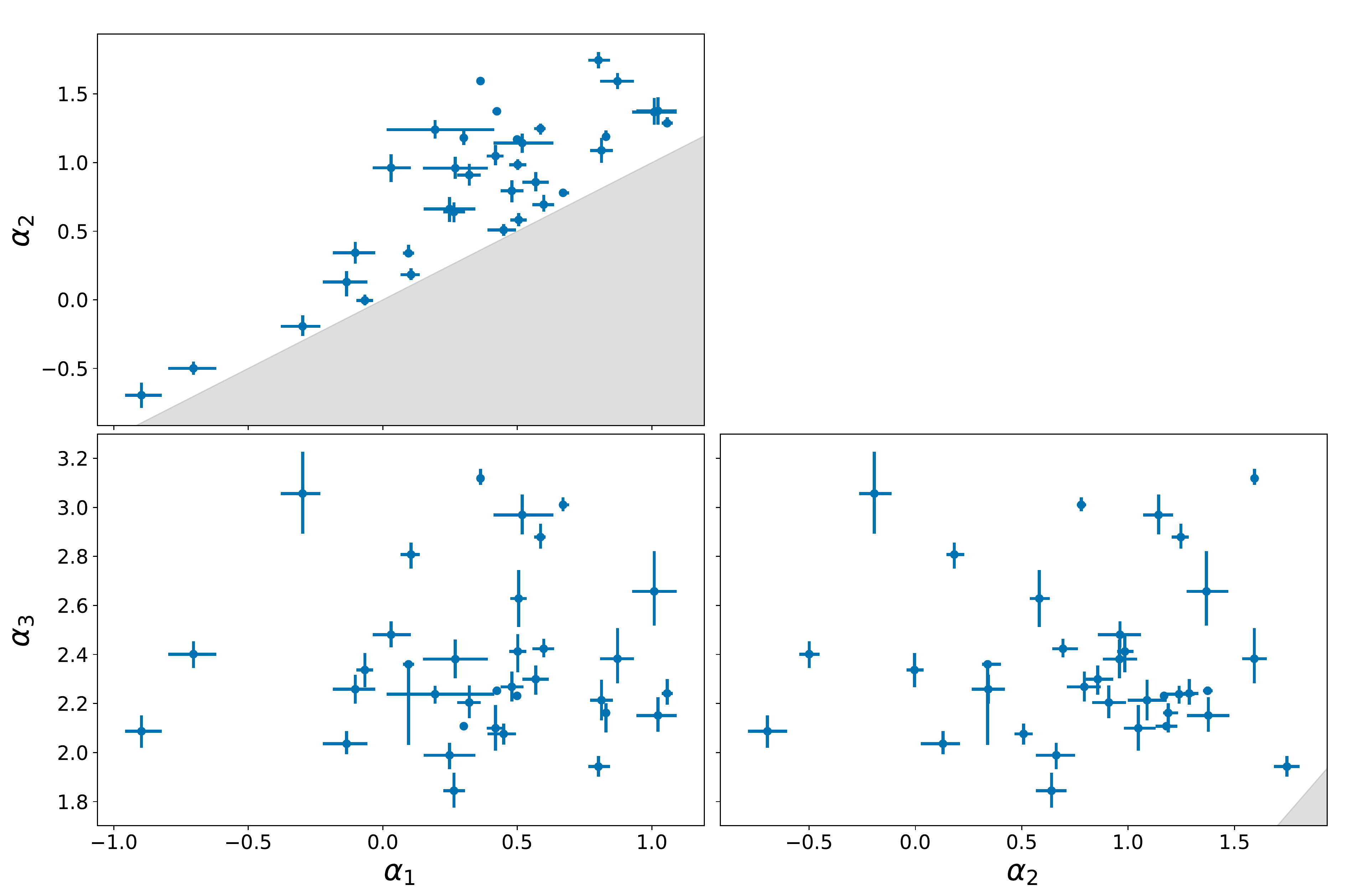}
        \caption{Relations between all three mass function exponent parameters.
                 Gray shaded areas represent the parameter space which is
                 disallowed by the priors on the mass function slopes.}
        \label{fig:ai_corner}
    \end{figure*}

    To begin, we examine the distribution of the \(\alpha\) parameters between
    all clusters.
    \Cref{fig:ai_corner} shows the relationships between all three mass function
    slopes (\(\alpha_1, \alpha_2, \alpha_3\)).  The high-mass \(\alpha_3\) is
    not clearly correlated to either of the other MF exponents, however a clear
    relation can be seen between \(\alpha_1\) and \(\alpha_2\). While it should
    be noted that, by design, the priors used here
    disallow \(\alpha_1 < \alpha_2\) (as shown by the shaded regions in
    \Cref{fig:ai_corner}), which may introduce a bias
    to this trend (although, as mentioned before, tests of clusters near this
    bound fit without this constraint resulted in negligible changes),
    it is clear that, in general, clusters with a more depleted
    low-mass mass function also have a relatedly depleted
    intermediate-mass mass function.

    The relation between \(\alpha_1\) and \(\alpha_2\) also showcases
    another important phenomenon; a large number of clusters do not fall on or
    near the \(\alpha_1 = \alpha_2\) line, but are instead steeper in the
    intermediate-mass regime than the low-mass regime. This suggests that
    a two-component power law is necessary to describe the global mass function
    below \(\SI{1}{\Msun}\), and a single power law attempting to describe the
    same mass regime would overestimate both the high and low mass ends of the
    mass function in this regime, and underestimate the mass function in the
    intermediate regime, near the break mass of \(\SI{0.5}{\Msun}\).
    That is, a single \(\alpha\) power law over the same mass domain would have
    a slope greater than \(\alpha_1\) and less than \(\alpha_2\).

\subsection{Initial mass function}


    The \(\alpha\) parameters constrained by the models describe the global
    stellar mass functions of the clusters at the present day, and in order to
    examine the initial mass function of our clusters we must carefully
    consider the connection between the IMF and the PDMF.
    As discussed in \Cref{sub:mass_function_evolution},
    we have chosen not to model the dynamical loss of (preferentially low-mass)
    stars in the mass function evolution algorithm used, due to its complex
    dependence on the dynamical evolution and initial conditions of the cluster.
    Therefore, in our models, the IMF can most directly be inferred only in
    the high-mass (\(\alpha_3, m> \SI{1}{\Msun}\)) regime, while the lower-mass
    exponents (\(\alpha_1, \alpha_2\)) are more representative of the
    present-day mass function, which may have evolved away from the IMF
    significantly.

    To quantify this assertion, we must examine the
    dynamical evolution of our clusters, as the dynamical loss of stars is not
    necessarily limited entirely to the lower-mass regime.
    In very dynamically evolved clusters,
    which have lost a substantial amount of their total initial mass to escaping
    stars, the characteristic mass of preferentially escaping stars will
    increase, potentially depleting even the population of higher-mass stars and
    WDs, which had initial masses above \SI{1}{\Msun}, and in such
    cases the inferred mass function exponent \(\alpha_3\) may also be shallower
    and less directly representative of the IMF. To account for this effect, we
    must determine which clusters have lost a large amount of their initial
    mass by the present day. We estimate this \textit{remaining mass fraction}
    by the equation:
    \begin{equation}\label{eq:remaining_mass_frac}
        \frac{M_{\mathrm{today}}}{M_{\mathrm{initial}}} = 
            0.55 \times \left(1- \frac{\mathrm{Age}}{\tau_{\mathrm{diss}}} \right)
    \end{equation}
    where the factor 0.55 reflects the typically assumed mass loss from
    stellar evolution of \(\sim 55 \%\) of the initial cluster mass in the
    first Gyr of a cluster's evolution and the dissolution time
    \(\tau_{\mathrm{diss}}\) represents the estimated total lifetime of the
    cluster. The estimated lifetimes of our
    clusters were computed according to the approach described in Section 3.2 of
    \citet{Baumgardt2019a}, using the updated models of \citet{Baumgardt2022}.
    This method is based on integrating the orbit of the clusters backwards
    in the Milky Way galactic potential \citep{Irrgang2013}, and
    estimating the resulting mass loss.
    A related quantity is the ``dynamical age'', which we define as
    the ratio of the cluster's age over its half-mass relaxation time
    (\(\tau_{\mathrm{rel}}\)).

    We have taken both the relaxation and dissolution times from the
    best-fitting models of \citet{Baumgardt2022}, a companion study which
    determined the mass functions of 120 MW, LMC and SMC globular clusters by
    comparing the same HST mass function datasets as used in this work with a
    grid of direct \Nbody simulations. While we could technically extract the
    relaxation times self-consistently from our own set of models, we utilize
    the values obtained by \citet{Baumgardt2022} in order to most easily
    compare our results. Given the good agreement (on average) between total mass and
    half-mass radii of their \Nbody models, as shown in \Cref{fig:model_comps},
    the differences should be negligible.

\begin{figure*}
        \centering
        \includegraphics[width=\linewidth]{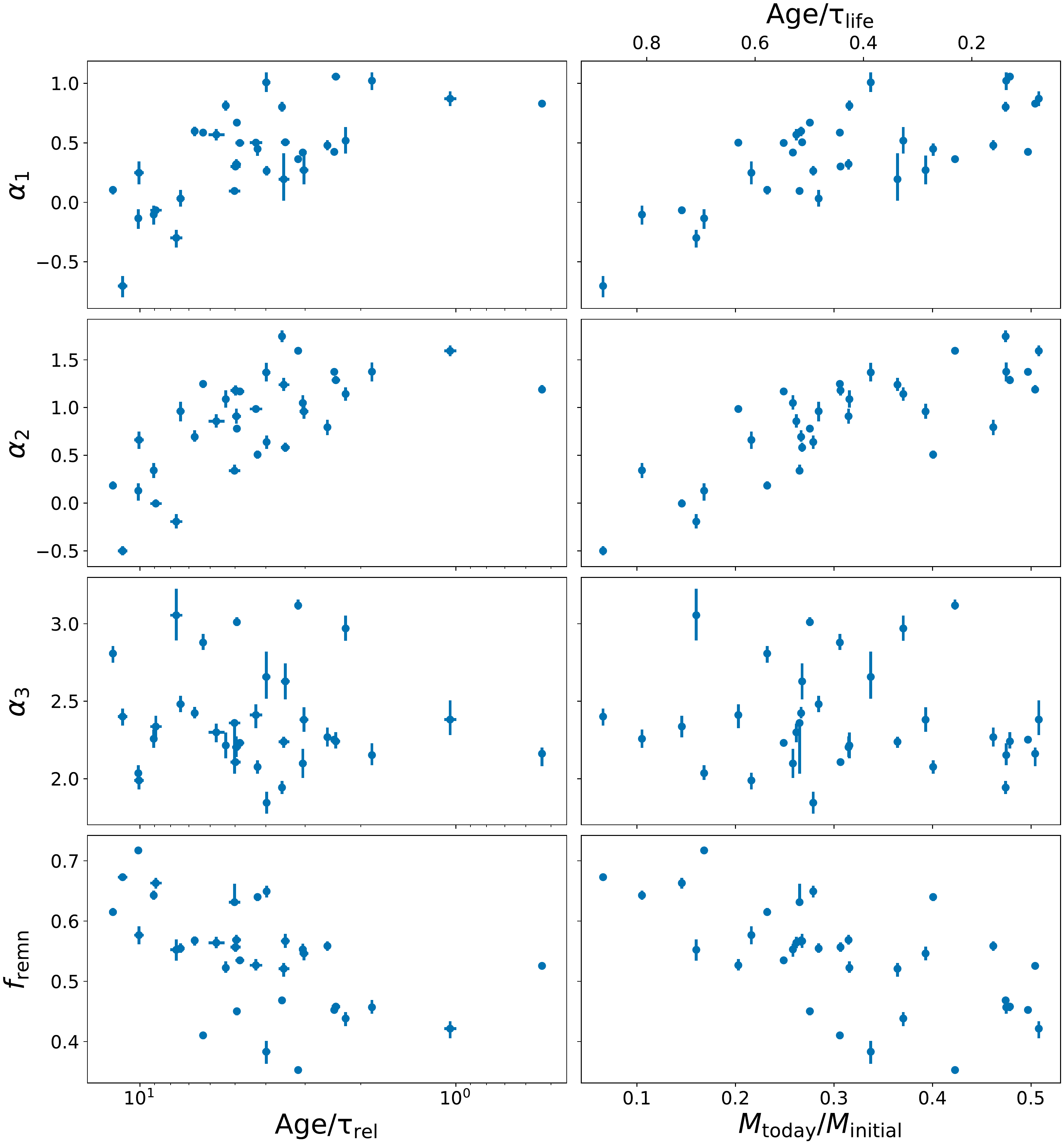}
        \caption{Relations between all three mass function exponent parameters
                 and the fraction of cluster mass in all stellar remnants (white
                 dwarfs, neutron stars, black holes)
                 versus the dynamical age and remaining mass fraction  of all
                 clusters. Clusters with higher remaining mass fractions
                 and relaxation times greater than their ages
                 provide more reliable probes of the IMF.}
        \label{fig:ai_mu_dyn}
    \end{figure*}

    These quantities, and their relationships with all mass function exponents,
    are shown for all clusters in \Cref{fig:ai_mu_dyn}. The
    clusters to the left of these plots are thought to have lost a large amount
    of their initial mass and be more dynamically evolved. In these cases the
    lower-mass \(\alpha_1\) and \(\alpha_2\) slopes are shallower (even
    becoming negative in the most dynamically evolved
    clusters), and the \(\alpha_3\) slope may also have
    been modified by dynamical evolution. As such, caution is advised when
    interpreting the slopes in these clusters as representative of the IMF.
    These quantities cannot be used to define an exact division of
    where the global mass function parameters reflect the IMF, but it does
    provide useful context to our proceeding analysis of the IMF.

    We can clearly see that both lower-mass MF exponents (\(\alpha_1\),
    \(\alpha_2\)) have distinct correlations with these two quantities, with
    the increasingly evolved clusters (short lifetimes / relaxation times
    compared to their ages) substantially more depleted in low-mass stars than
    their less evolved counterparts.
    This trend, and the IMF in the low and intermediate-mass regime are explored
    in more detail in \citet{Baumgardt2022}.
    No such correlation exists with \(\alpha_3\), which supports
    our assertion that the high-mass regime is less affected by the cluster's
    dynamical evolution, and thus, overall, most representative of the IMF.
    However, as stated before, caution should still be applied when interpreting
    the \(\alpha_3\) of the clusters to the left side of this figure.
    We will examine this parameter in more detail in \Cref{sub:high_mass_imf}
    below.

    The evolution of the remnant mass fraction \(f_{\mathrm{remn}}\), which
    includes all types of stellar remnants, is also shown at the bottom of
    \Cref{fig:ai_mu_dyn}, where a strong relationship with the dynamical age
    of the clusters is evident, as
    might be expected. As a cluster evolves and loses mass, as mentioned
    before, the mass lost is preferentially in the form of lower-mass
    stars, rather than the heavy remnants, and as such the fraction of mass in
    remnants should increase as the cluster's low-mass MF is depleted.
    Interestingly, some of the most dynamically evolved clusters have nearly
    75\% of their mass in dark remnants at the present day, which could have
    important implications for the mass-to-light ratios and inferred masses of
    unresolved GCs in distant galaxies.

\subsubsection{High-mass IMF}\label{sub:high_mass_imf}


    \begin{figure*}
        \centering
        \includegraphics[width=\linewidth]{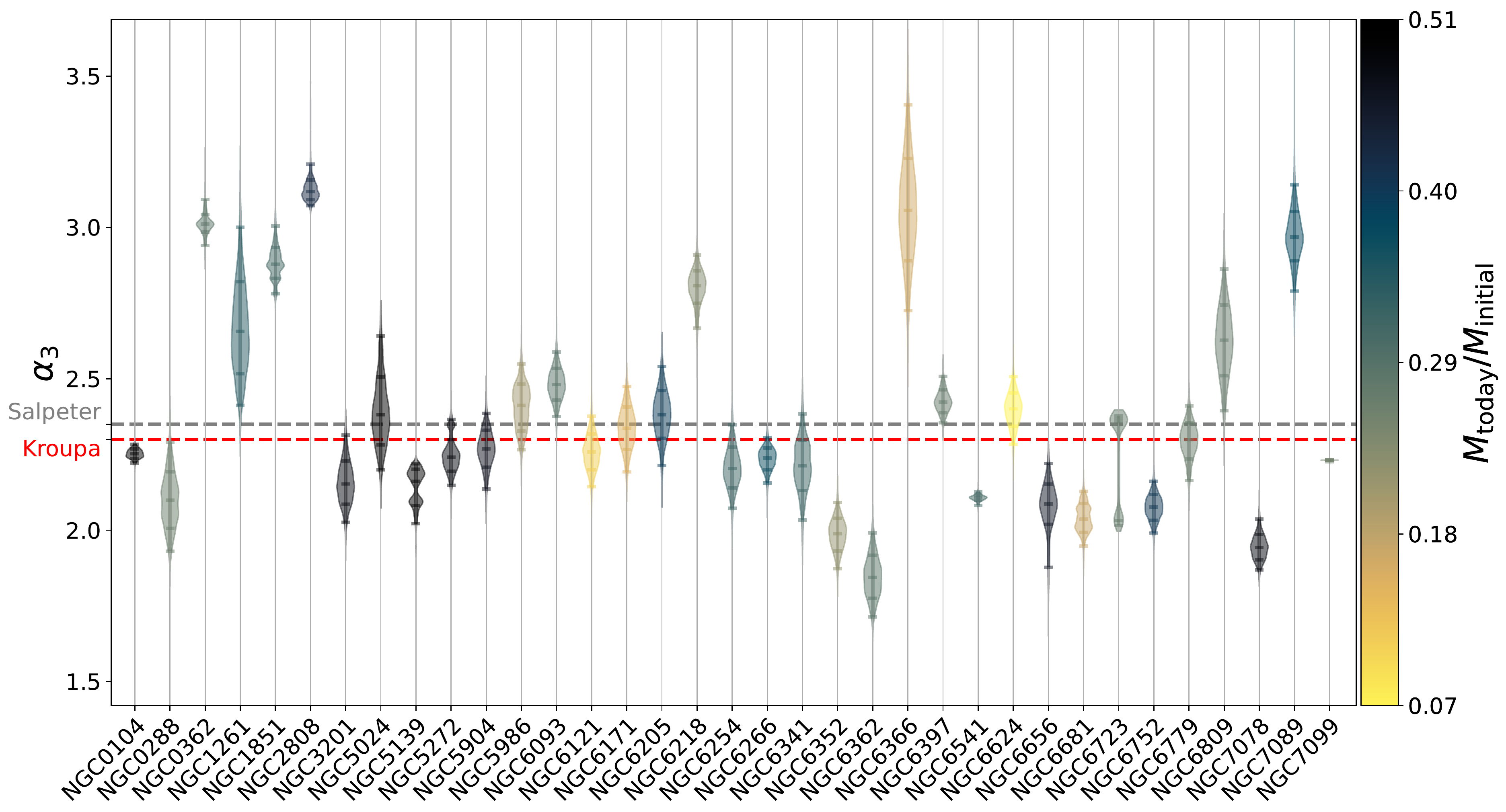}
        \caption{Violin plot of the \(\alpha_3\) parameter posterior
                 distributions for all clusters.
                 The median, \(1\sigma\) and \(2\sigma\) values are denoted by
                 the horizontal ticks on each distribution.
                 Colours represent the remaining mass fraction. The
                 corresponding values of some canonical
                 high-mass (\(m>\SI{1}{\Msun}\)) IMF formulations
                 (\citealp{Salpeter1955,Kroupa2001}) are shown by dashed lines.}
        \label{fig:a3_distributions}
    \end{figure*}

    \Cref{fig:a3_distributions} shows the posterior probability distributions
    of \(\alpha_3\) for all clusters.
    From this figure we can see that the distributions are, in the vast majority
    of cases, compatible within uncertainties with the typically
    assumed canonical high-mass (\(m>\SI{1}{\Msun}\)) IMF formulations
    \citep[e.g.][]{Salpeter1955,Kroupa2001}, however with a large spread of
    \(\alpha_3\) values between \(\sim 2\mathrm{-}3\). The median and
    \(1\sigma\) values over all clusters are
    \(\alpha_3 = 2.37\substack{+0.48 \\ -0.25}\).
    This matches remarkably well with the canonical IMFs, a striking result
    given the large freedom in the mass function of our models.
    This result is also in agreement with the high-mass slopes determined
    by \citet{Baumgardt2022} through the examination of similar HST
    mass function datasets in younger clusters in the Large (LMC) and Small
    Magellanic Clouds (SMC), where more massive stars, yet to evolve off the
    main sequence, can still be observed directly.
    Similar results were also obtained by \citet{Weisz2015} for young clusters
    in \Messier{31}.
    It is even clearer that our fits \textit{do not} favour any more extreme
    IMFs, neither exceedingly top-heavy nor top-light, especially
    when ignoring the most dynamically evolved clusters (shown in
    \Cref{fig:a3_distributions} by the more yellow colours).

    This result is counter to some recent suggestions in the literature of
    top-heavy IMFs
    in GCs. It has been shown that clusters with top-heavy IMFs are expected to
    have lost a very large fraction of their mass early in their lifetimes due
    to stellar mass loss and supernova explosions \citep{Haghi2020},
    produce a large amount of BHs and could contribute significantly to the
    observed rate of binary BH mergers \citep{weatherford2021,Antonini2022}.
    Given that our results seem to preclude any clusters as top-heavy as
    \(\alpha_3\sim1.6\), there is thus no obvious need to consider top-heavy
    IMFs in estimates of BBH merger rates in globular clusters.
    Due to the smaller dissolution times of top-heavy GCs of typical
    masses (\(\sim 10^5\ \Msun\)), there remains the possibility that some
    GCs had formed with a more top-heavy IMF, and have simply dissolved to such
    an extent by the present day that they are undetectable. These clusters
    could still contribute significantly to the rate of BBH mergers and
    gravitational waves. However, given the large range of GC parameter space
    covered by our models, it is unclear what would cause these top-heavy
    GCs to form alongside clusters with a more canonical IMF as we see here.
    As shown in \citet{Haghi2020}, the dissolution times of GCs scale
    smoothly with the IMF, resulting in, for example, lifetimes \(\sim 3\)
    times shorter for typical-mass clusters with \(\alpha_3 \approx 1.8\),
    compared to clusters with a canonical IMF. Given that the spread in initial
    masses of the GC population is likely on the order of \(\sim 100\)
    \citep[e.g.][]{Balbinot2018}, we would expect to still find some surviving
    clusters with an \(\alpha_3 < 1.8\), if they had formed alongside our
    sample.

    It should be noted again that, as mentioned in \Cref{sec:cluster_parameters},
    the uncertainties on these parameters represent only the statistical
    uncertainties on the fits, and the actual errors could be larger.
    This extra uncertainty on the model would be difficult to quantify exactly,
    however, examining in particular \(\alpha_3\), based on a reduced
    chi-squared test, if all the scatter in these results (away from the mean)
    were explained by the uncertainties in our models (beyond the statistical
    uncertainties shown), it would indicate a typical error
    of approximately 0.4 on \(\alpha_3\).

\subsection{Relationship with metallicity}


    We next examine possible correlations between the high-mass stellar IMF
    of GCs and metallicity.
    Variations of the initial mass function with metallicity have been suggested
    in the past based on theoretical studies of star and cluster
    formation, which indicate that increasing metallicity leads to more
    efficient cooling and helps limit stellar accretion, and thus should
    reduce the characteristic mass of formed stars and produce an increasingly
    bottom-heavy IMF in more metal-rich clusters
    \citep{Larson1998,Demarchi2017,Chon2021}.
    \citet{Marks2013} proposed a linear relationship between the high-mass
    IMF slope and metallicity, which begins with more top-heavy values of
    \(\alpha_3\) at lower metallicities (\(\alpha_3 = 1\) at \(\FeH = -2.5\)),
    and reaches the canonical Kroupa value of 2.3 only at metallicities
    \(\FeH > -0.5\).
    Given the large amount of freedom available in our mass function slopes,
    and the excellent constraints we are able to place on the dark remnant
    populations in this mass regime, our model fits, which span nearly the full
    range of Milky Way GC metallicities, present an excellent opportunity
    to examine this potential relationship.

    \begin{figure*}
        \centering
        \includegraphics[width=\linewidth]{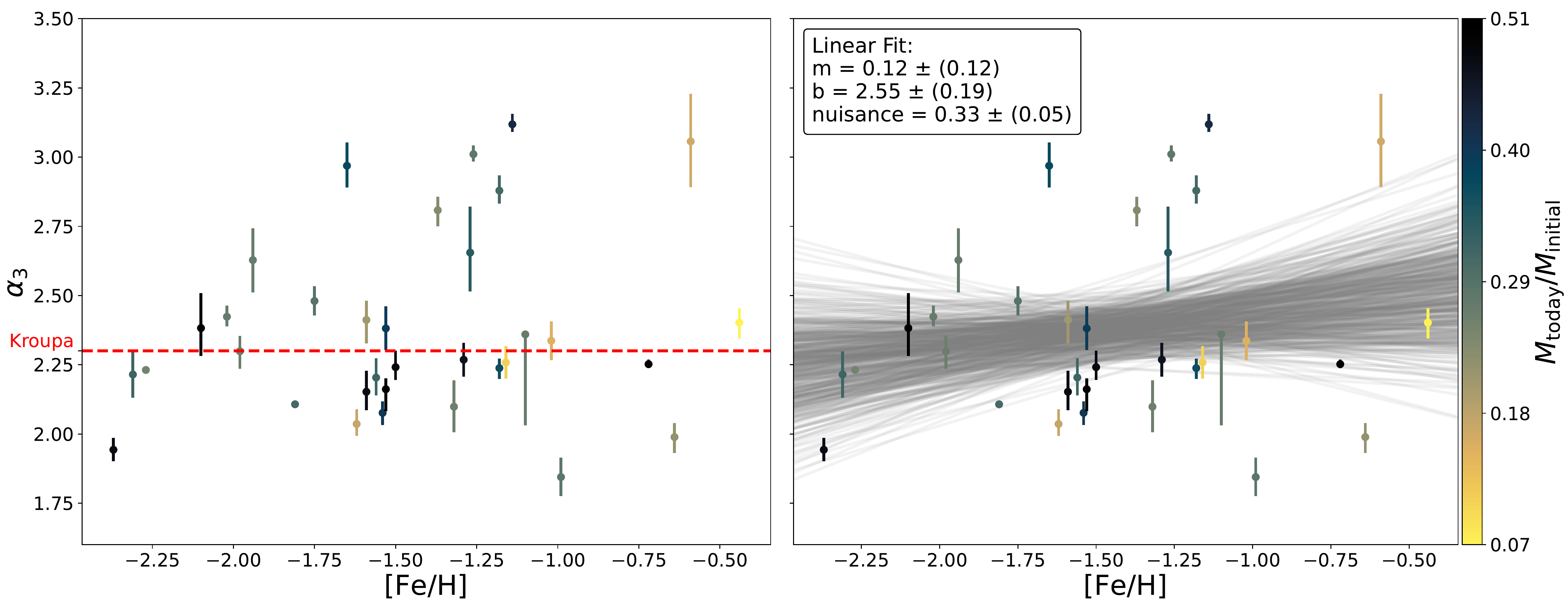}
        \caption{Relation between the high-mass IMF exponent \(\alpha_3\) and
                 the cluster metallicity for all clusters.
                 Colours represent the remaining mass fraction.
                 The corresponding value of the \citet{Kroupa2001} canonical
                 high-mass (\(m>\SI{1}{\Msun}\)) IMF formulation
                 is shown by the red dashed line.
                 On the right panel, over-plotted in grey is the best-fit linear
                 relation representing the given clusters, shown by 500 random
                 draws of the converged MCMC chain. The median and \(1\sigma\)
                 uncertainties of the parameters of these fits are given in the
                 upper-left corner of each panel.}
        \label{fig:a3_FeH}
    \end{figure*}

    \Cref{fig:a3_FeH} shows the relationship between \(\alpha_3\) and cluster
    metallicity \FeH. As can be seen in the left panel, while there does seem to
    be an absence of more top-light clusters at lower metallicities, no clear
    overall trend or relationship seems to emerge. Most clusters are, as stated
    before, scattered around the canonical \(\alpha_3\) value, with a large
    spread but no apparent dependence on metallicity.
    %

    To probe for any potential trend further, we attempt to directly fit a
    linear relation (\(\alpha_3 = m \times \FeH + b\)) to this
    plot, using a simple MCMC sampler with a Gaussian likelihood (using
    \texttt{emcee}; \citealp{Foreman-Mackey2016}). To account
    for any biases and underestimated uncertainties in our inferred \(\alpha_3\)
    values (as discussed in \Cref{sec:cluster_parameters,sub:high_mass_imf}),
    we also include a nuisance parameter, added in quadrature to the
    statistical errors.
    The right panel of \Cref{fig:a3_FeH} shows the results of this
    fit. The linear fit shows a very slightly positive median slope, but
    within uncertainties is entirely consistent with no correlation at all.
    The best-fit nuisance parameter value is also remarkably
    similar to the estimated model uncertainties computed in
    \Cref{sub:high_mass_imf}.

    This analysis is somewhat limited by the
    smaller number of clusters with a larger remaining mass fraction at both
    extremes of the metallicity range of Milky Way GCs, which largely drive
    the results of this fit. Further extension
    of this work to more metal-poor and metal-rich clusters would aid in
    definitively supporting or excluding the existence of a correlation
    between the metallicity and the stellar IMF of GCs.

    It is clear though, even with these caveats, that the very top-heavy
    IMF-metallicity relationship proposed by \citet{Marks2013} is not compatible
    with our results. As discussed in \Cref{sub:high_mass_imf}, none of our
    clusters favour a top-heavy IMF, and even our most metal-poor clusters have
    a value of \(\alpha_3\) much closer to the canonical \(\sim2.3\) than
    suggested by the fundamental plane of \citet{Marks2013}.



\section{Conclusions}\label{ch:conclusions}

    %
    In this work, we have inferred, through dynamic nested sampling,
    the best-fitting model parameter distributions of multimass
    \limepy models for a large sample of Milky Way globular clusters, subject
    to a number of observed proper motion, line-of-sight velocity, number
    density and stellar mass function datasets.
    %
    %
    This process has resulted in well-fit models for 34 Milky Way GCs, with
    full, well constrained, posterior distributions for the structural, mass
    functions, and helicocentric distance parameters of each cluster.
    These results show excellent matches with the properties of the \Nbody
    models computed by \citet{Baumgardt2019a}.

    These models further allow us to explore in detail the stellar (initial)
    mass functions of a large sample of Milky Way GCs, and
    yield a number of important conclusions:

    \begin{enumerate}


        \item
        Deviations of the low and intermediate-mass stellar mass function slopes
        from the \(\alpha_1 = \alpha_2\) line demonstrate that a
        two-component power law is necessary in order to describe the (initial)
        mass function in this mass regime.

        \item
        We show that, while the low and intermediate-mass MF slopes are
        strongly dependent on the dynamical age of the clusters, the high-mass
        slopes (\(\alpha_3; m > \SI{1}{\Msun}\)) are not, indicating that the
        MF in this regime has generally been less affected by dynamical losses,
        and is most representative of the IMF.

        \item
        Examination of the high-mass MF slopes suggest an IMF in this regime
        (\(\alpha_3 = 2.37\substack{+0.48 \\ -0.25}\))
        which is in excellent agreement with canonical values
        \citep[e.g.][]{Salpeter1955,Kroupa2001}. This result precludes the need
        for any more extreme high-mass IMF formulation for globular clusters,
        such as a top-heavy IMF.

        \item
        No statistically significant correlation is found between the high-mass
        stellar IMF slope in GCs and cluster metallicity.

    \end{enumerate}

    In a separate paper (\paperII), we will use the best-fitting models
    presented in this work to analyze and discuss the populations of
    stellar-mass BHs in our sample of GCs.

\section*{Acknowledgements}

ND is grateful for the support of the Durland Scholarship in Graduate Research.
VHB acknowledges the support of the Natural Sciences and Engineering Research Council of Canada (NSERC) through grant RGPIN-2020-05990.
MG acknowledges support from the Ministry of Science and Innovation (EUR2020-112157, PID2021-125485NB-C22, CEX2019-000918-M funded by MCIN/AEI/10.13039/501100011033) and from AGAUR (SGR-2021-01069).

This research was enabled in part by support provided by ACENET (\url{www.ace-net.ca}) and the Digital Research Alliance of Canada (\url{https://alliancecan.ca}).

This work has also benefited from a variety of \texttt{Python}
packages including
\texttt{astropy} \citep{Astropy2013,Astropy2018},
\texttt{corner} \citep{Foreman-Mackey2016},
\texttt{dynesty} \citep{Speagle2020},
\texttt{emcee} \citep{Foreman-Mackey2016},
\texttt{h5py} \citep{Collette2022},
\texttt{matplotlib} \citep{Hunter2007},
\texttt{numpy} \citep{Harris2020},
\texttt{scipy} \citep{Virtanen2020} and
\texttt{shapely} \citep{Gillies2022}.

\section*{Data Availability}

The data underlying this article are available at
\url{https://github.com/nmdickson/GCfit-results}.
Extracted Gaia EDR3 PM dispersion profiles are also available in Zenodo, at
\url{https://dx.doi.org/10.5281/zenodo.7344596}.


\bibliographystyle{mnras}
\bibliography{biblio}


\appendix

\section{Data sources for each cluster}\label{appendix:data_sources}

\begin{table}
    \begin{threeparttable}[b]
    \caption{The literature sources of all number density (ND), line-of-sight
             velocity dispersion (LOS), proper motion velocity dispersion (PM)
             and mass function (MF) datasets, for each cluster in our sample.}
    \renewcommand*{\arraystretch}{1.1}
    \centering
    \begin{tabular}{rllll}
        \hline
        Cluster  &  \multicolumn{1}{c}{ND} & \multicolumn{1}{c}{LOS}  & \multicolumn{1}{c}{PM}  & \multicolumn{1}{c}{MF} \\
        \hline
        \NGC104  & dB19 & B18, K18      & L22, GEDR3      & B23 \\
        \NGC288  & dB19 & B18           & L22, GEDR3      & B23 \\
        \NGC362  & dB19 & B18, K18      & L22, GEDR3      & B23 \\
        \NGC1261 & dB19 & B18           & L22, GEDR3      & B23 \\
        \NGC1851 & dB19 & B18, K18, L13 & L22, GEDR3      & B23 \\
        \NGC2808 & dB19 & B18, K18      & L22, GEDR3      & B23 \\
        \NGC3201 & dB19 & B18           & L22, GEDR3      & B23 \\
        \NGC4590 & dB19 & B18           & L22, GEDR3      & B23 \\
        \NGC5024 & dB19 & B18           & L22, GEDR3      & B23 \\
        \NGC5139 & dB19 & B18, K18      & W15, GEDR3      & B23 \\
        \NGC5272 & dB19 & B18           & L22, GEDR3      & B23 \\
        \NGC5904 & dB19 & B18, K18      & L22, GEDR3      & B23 \\
        \NGC5986 & dB19 & B18           & L22, GEDR3      & B23 \\
        \NGC6093 & dB19 & B18, K18, L13 & L22, GEDR3      & B23 \\
        \NGC6121 & dB19 & B18           & L22, GEDR3      & B23 \\
        \NGC6171 & dB19 & B18           & L22, GEDR3      & B23 \\
        \NGC6205 & dB19 & B18           & L22, GEDR3      & B23 \\
        \NGC6218 & dB19 & B18           & L22, GEDR3      & B23 \\
        \NGC6254 & dB19 & B18, K18      & L22, GEDR3      & B23 \\
        \NGC6266 & dB19 & B18, K18, L13 & W15, GEDR3      & B23 \\
        \NGC6341 & dB19 & B18           & L22, GEDR3      & B23 \\
        \NGC6352 & dB19 & B18           & L22, GEDR3      & B23 \\
        \NGC6362 & dB19 & B18,D21       & L22, GEDR3      & B23 \\
        \NGC6366 & dB19 & B18           & L22, GEDR3      & B23 \\
        \NGC6397 & dB19 & B18           & L22, GEDR3      & B23 \\
        \NGC6541 & dB19 & B18, K18      & L22, GEDR3      & B23 \\
        \NGC6624 & dB19 & B18           & L22, GEDR3      & B23 \\
        \NGC6656 & dB19 & B18           & L22, GEDR3      & B23 \\
        \NGC6681 & dB19 & B18, K18      & L22, GEDR3      & B23 \\
        \NGC6723 & dB19 & B18           & L22, GEDR3, T22 & B23 \\
        \NGC6752 & dB19 & B18           & L22, GEDR3      & B23 \\
        \NGC6779 & dB19 & B18           & L22, GEDR3      & B23 \\
        \NGC6809 & dB19 & B18           & L22, GEDR3      & B23 \\
        \NGC6981 & dB19 & B18           & L22, GEDR3      & B23 \\
        \NGC7078 & dB19 & B18, K18      & L22, GEDR3      & B23 \\
        \NGC7089 & dB19 & B18, K18      & L22, GEDR3      & B23 \\
        \NGC7099 & dB19 & B18, K18      & L22, GEDR3      & B23 \\
        \hline
    \end{tabular}
    \begin{tablenotes}
        \item [dB19] \citep{deBoer2019}
        \item [B18] \citep{Baumgardt2017,Baumgardt2018}
        \item [K18] \citep{Kamann2018}
        \item [D21] \citep{Dalessandro2021}
        \item [L13] \citep{Lutzgendorf2013}
        \item [L22] \citep{Libralato2022}
        \item [W15] \citep{Watkins2015}
        \item [T22] \citep{Taheri2022}
        \item [GEDR3] This work (Gaia EDR3)
        \item [B23] \citep{Baumgardt2022}
    \end{tablenotes}
    \end{threeparttable}
\end{table}


\bsp	
\label{lastpage}
\end{document}